\documentclass[preprint,amsmath,amssymb,aps,nofootinbib]{revtex4-2}

\usepackage{graphicx}
\usepackage{caption}
\usepackage{subcaption}
\usepackage{dcolumn}
\usepackage{bm}
\usepackage{physics}
\usepackage{xcolor}
\usepackage{filecontents}
\usepackage{soul}

\allowdisplaybreaks

\begin{document}

\title{Second-order stochastic theory for self-interacting scalar fields in de Sitter spacetime}
\author{Archie Cable}
 \email{archie.cable18@imperial.ac.uk}
\author{Arttu Rajantie}
 \email{a.rajantie@imperial.ac.uk}
\affiliation{Department of Physics,\\Imperial College London,\\London, SW7 2AZ, United Kingdom}

\date{\today}

\begin{abstract}
    We introduce a second-order stochastic effective theory for light scalar fields in de Sitter spacetime, extending the validity of the stochastic approach beyond the massless limit and demonstrating how it can be used to compute long-distance correlation functions non-perturbatively. The parameters of the second-order stochastic theory are determined from quantum field theory through a perturbative calculation, which is valid if the self-interaction parameter $\lambda$ satisfies $\lambda\ll m^2/H^2$, where $m$ is the scalar and $H$ is the Hubble rate. Therefore it allows stronger self-interactions than conventional perturbation theory, which is limited to $\lambda\ll m^4/H^4$ by infrared divergences. We demonstrate the applicability of the second-order stochastic theory by comparing its results with perturbative quantum field theory and overdamped stochastic calculations, and discuss the prospects of improving its accuracy with a full one-loop calculation of its parameters.
\end{abstract}

\maketitle

\tableofcontents

\section{Introduction}
\label{sec:intro}

Scalar quantum field theory (QFT) in de Sitter spacetime is a widely studied topic \cite{Birrell-Davies_book,bunch-davies:1978,tagirov:1973,chernikov:1968}, especially in the context of inflationary cosmology \cite{Starobinsky:1980,guth_inflation,linde_inflation,slow_roll_liddle,Vazquez_Gonzalez_2020_slowroll}. Of particular interest is the study of the long distance behaviour of spectator scalar fields as they can lead to present-day observables with blueprints from inflation \cite{baumann_book}. Physical examples include curvature and isocurvature perturbations \cite{Sasaki:1986, Linde:1997, Leach:2001}, dark matter generation, \cite{Peebles_1999_darkmatter, Hu_2000_darkmatter, Markkanen_2018_darkmatter,Jukko:2021}, electroweak vacuum decay \cite{Espinosa_2008_vdecay,Herranen_2014_vdecay,Markkanen_2018_vdecay,Camargo-Molina:2022p1,Camargo-Molina:2022p2} and gravitational wave background anisotropies \cite{gwb_anisotropy}.

The issue with scalar QFT in de Sitter is that self-interactions, parameterised by the coupling $\lambda$, lead to the existence of infrared divergences in perturbation theory that cannot be dealt with using standard methods beyond the limit $\lambda\ll m^4/H^4$ where $m$ and $H$ are the scalar mass and Hubble rate respectively \cite{allen_folacci_perturbative_corr, allen_perturbative_corr, Sasaki:1993, Suzuki:1994}. This is problematic when studying the long distance behaviour of these fields, and that has led physicists to consider alternative approaches, some of which do not require small $\lambda$ but involve other approximations
\cite{hu_oconnor_symm_behaviour,boyanovsky_quantum_correct_SR,Serreau_2011,Arai_2012,Gautier_2013,Herranen_2014,Gautier_2015,Guilleux_2015,Boyanovsky:2015,Guilleux:2016,Boyanovsky:2016,Nacir:2019}. 
One such method is the stochastic approach \cite{starobinsky:1986,Starobinsky-Yokoyama:1994}, an effective theory where one utilises the fact that the expansion of the inflationary spacetime causes long-wavelength modes to be stretched across the de Sitter horizon such that they can be considered classical. The remaining short-wavelength quantum modes then contribute in the form of stochastic noise. This will only be possible if $m$ is comparable to or smaller than the horizon scale, $m\lesssim H$.

The most common application of this method is to stochastic inflation where the scalar field plays the part of the inflaton \cite{ Tsamis_2005,Finelli_2009,Finelli_2010,Vennin_2015,Hardwick_2017_post-inflation,Tokuda_2018,Tokuda_2018_2,Glavan_2018,Cruces_2019,Grain_2017,Firouzjahi_2019,Pinol_2019,Hardwick_2019,Pattison_2019,Moreau_2020,Moreau:2020gib,Rigopoulos_2016,Moss_2017,Prokopec_2018,bounakis2020feynman,garbrecht_rigopoulos:2014,Garbrecht_2015_Fdiag,morikawa_1990,rigopoulos2013fluctuationdissipation,Levasseur_2013,Rigopoulos_2016,Moss_2017,Pinol_2019, Pinol:2020,Moreau_2020,Moreau:2020gib,Wilkins:2021,Andersen_2022}. In this case, the fields are considered to be in slow-roll such that one can use the overdamped approximation in the stochastic equations, which equates to neglecting the second derivative of the field. For this to be valid, we require the fields to be light $m\ll H$ and for the coupling to be sufficiently small $\lambda\ll m^2/H^2$. In this limit, one can derive the overdamped stochastic equations by introducing a cut-off between sub- and superhorizon field modes with a strict boundary \cite{starobinsky:1986,Starobinsky-Yokoyama:1994}. The superhorizon modes then play the role of a classical field while the subhorizon ``quantum'' modes contribute via a stochastic white noise. This mode expansion derivation has been widely studied \cite{Tsamis_2005,Finelli_2009,Finelli_2010,Vennin_2015,Hardwick_2017_post-inflation,Tokuda_2018,Tokuda_2018_2,Glavan_2018,Cruces_2019,Grain_2017,Firouzjahi_2019,Markkanen:2019,Pinol_2019,Hardwick_2019,Pattison_2019,Markkanen_2020,Moreau_2020,Moreau:2020gib} alongside an alternative path integral approach that aligns more with standard thermal field theory methods \cite{Rigopoulos_2016,Moss_2017,Prokopec_2018,bounakis2020feynman,garbrecht_rigopoulos:2014,Garbrecht_2015_Fdiag,morikawa_1990,rigopoulos2013fluctuationdissipation,Levasseur_2013, Pinol_2019, Pinol:2020,Wilkins:2021,Andersen_2022}. 

The stochastic approach has not exclusively been applied to the inflaton. From the seminal work of Starobinsky and Yokoyama \cite{starobinsky:1986,Starobinsky-Yokoyama:1994} to more recent applications to inflationary observables \cite{Markkanen_2018_darkmatter,Markkanen:2019,Markkanen_2020,Jukko:2021,Cable:2021}, there are examples of the stochastic approach being applied to spectator scalar fields. In this case, the slow-roll condition can be relaxed and one should consider the full second-order stochastic equations. However, in our previous paper \cite{Cable:2021}, we showed that for free fields the standard cut-off procedure cannot be used to derive the second-order stochastic equations when one goes beyond the limit $m\ll H$. This is not a fault in the stochastic equations themselves but in the expression for the noise amplitude that one computes using a cut-off. Given the correct form of the noise, the quantum correlators can be be recovered from these second-order equations. Following the standard technique for effective field theories, this form was found by evaluating the stochastic field correlator for an unspecified noise amplitude and then matching with the quantum field propagator in order to determine the desired noise amplitude. This suggests that the stochastic approach is still viable away from the overdamped limit but that the strict cut-off procedure is no longer a valid means of deriving the stochastic equations from the underlying quantum theory.

In this paper, we take this idea a step further and introduce interactions into the stochastic equations. We relax the overdamped approximation that $m\ll H$ to show that in perturbation theory, we can once again use the matching procedure to find the form of the stochastic parameters, namely the mass, coupling and noise amplitudes, that reproduces the quantum correlators. Further, we present a numerical method for solving the second-order stochastic equations such that stochastic correlation functions can be found non-perturbatively.

The results we obtain using the second-order stochastic effective theory in this work are valid for fields with mass $m\lesssim H$ and quartic self-interaction coupling $\lambda\ll m^2/H^2$. We leave it to future work \cite{in_preparation} to compute $\mathcal{O}\qty(\frac{\lambda H^2}{m^2})$ contributions, where we will have to employ a renormalisaton scheme in the QFT in order to compute the stochastic noise amplitudes. However, we are already going beyond perturbative techniques in QFT and the overdamped approximation, which are only valid in the regimes $\lambda\ll m^4/H^4$ and $m\ll H$ respectively. The second-order stochastic approach encompasses these regions and goes further where it can compute physical results that the established approximations cannot. 

The paper is organised in the following way. In Sec. \ref{sec:qft}, we review the status of perturbative QFT and discuss the arising issues, focussing on a self-interacting theory with quartic coupling. In Sec. \ref{sec:stochastic_approach}, we briefly summarise the overdamped stochastic approach before introducing the second-order stochastic equations. We introduce a spectral expansion method to compute stochastic correlators and give the form of these in terms of the eigenspectrum. In Sec. \ref{sec:comparison_with_perturbative_QFT}, we begin by summarising the results from Ref. \cite{Cable:2021} for free fields before using perturbation theory to compute stochastic correlators to leading order in the self-interaction coupling. We compare these results to their equivalents at $\mathcal{O}\qty(\frac{\lambda H^4}{m^4})$ in perturbative QFT, from which we find the value of the stochastic parameters required to reproduce the quantum result. In Sec. \ref{sec:numerical_calculation}, we outline the numerical method to evaluate correlators non-perturbatively and then, using the noise functions found in Sec. \ref{sec:comparison_with_perturbative_QFT}, we perform a full comparison between the second-order stochastic, overdamped stochastic and perturbative QFT approaches. Finally, we discuss the results and conclude in Sec. \ref{sec:discussion}.

\section{Quantum field theory in de Sitter spacetime}
\label{sec:qft}

We will begin by reviewing the status of scalar QFT in de Sitter, focussing on the calculation of the Feynman propagator. We will consider a spectator scalar field $\phi(t,\mathbf{x})$ with scalar potential $V(\phi)$ on a de Sitter background with scale factor $a(t)=e^{Ht}$. $H=\Dot{a}/a$ is the Hubble rate which will be kept constant throughout. Introducing the field momentum $\pi(t,\mathbf{x})$, we can write the equations of motion as

\begin{equation}
    \label{scalar_eom}
    \begin{pmatrix}\Dot{\phi}\\\Dot{\pi}\end{pmatrix}=\begin{pmatrix}\pi\\-3H\pi-V'(\phi)\end{pmatrix},
\end{equation}

\noindent where primes and dots denote derivatives with respect to $\phi$ and $t$ respectively. We will focus on a $\phi^4$ theory such that the potential $V(\phi)=\frac{1}{2}m^2\phi^2+\frac{1}{4}\lambda\phi^4$ where $m$ is the scalar mass and $\lambda$ is the quartic coupling constant. We introduce a non-minimal coupling to gravity $\xi$, which is included in the scalar mass term as $m^2=m_0^2+12\xi H^2$.

\subsection{Free quantum fields}

For free fields ($\lambda=0$), one can follow standard QFT procedures to calculate the Feynman propagator, resulting in \cite{Birrell-Davies_book,tagirov:1973, chernikov:1968, bunch-davies:1978}

\begin{equation}
    \label{qft_feynman_propagator}
    \begin{split}
    i\Delta(t,t',\mathbf{x},\mathbf{x}'):=&\expval{\hat{T}\hat{\phi}(t,\mathbf{x})\hat{\phi}(t',\mathbf{x}')}\\=&\frac{H^2}{16\pi^2}\Gamma\qty(\frac{3}{2}+\nu)\Gamma\qty(\frac{3}{2}-\nu){_2}F_1\qty(\frac{3}{2}+\nu,\frac{3}{2}-\nu,2;1+\frac{y}{2})
    \end{split}
\end{equation}

\noindent in the Bunch-Davies vacuum, where $_2F_1(a,b,c;z)$ is the hypergeometric function, $\Gamma(z)$ are the Euler-Gamma functions, $\nu=\sqrt{\frac{9}{4}-\frac{m^2}{H^2}}$ and $y$ is the spacetime interval given by

\begin{equation}
    \label{de_sitter_inv_quantity}
    y=\cosh(H(t-t'))-\frac{H^2}{2}a(t)a(t')\abs{\mathbf{x}-\mathbf{x}'}^2-1.
\end{equation}

\noindent We can write the Feynman propagator instead as

\begin{equation}
    \label{asymptotic_sum_QFT_corr}
    \begin{split}
    i\Delta(t,t',\mathbf{x},\mathbf{x}')=&\frac{H^2}{16\pi^2}\Bigg[\frac{\Gamma(2\nu)\Gamma(1-2\nu)}{\Gamma\qty(\frac{1}{2}+\nu)\Gamma\qty(\frac{1}{2}-\nu)}\sum_{n=0}^{\infty}\frac{\Gamma\qty(\frac{3}{2}-\nu+n)\Gamma\qty(\frac{1}{2}-\nu+n)}{\Gamma\qty(1-2\nu+n)n!}\qty(-\frac{y}{2})^{-\frac{3}{2}+\nu-n}\\&\\
    &+\frac{\Gamma(-2\nu)\Gamma(1+2\nu)}{\Gamma\qty(\frac{1}{2}+\nu)\Gamma\qty(\frac{1}{2}-\nu)}\sum_{n=0}^{\infty}\frac{\Gamma\qty(\frac{3}{2}+\nu+n)\Gamma\qty(\frac{1}{2}+\nu+n)}{\Gamma\qty(1+2\nu+n)n!}\qty(-\frac{y}{2})^{-\frac{3}{2}-\nu-n}\Bigg],
    \end{split}
\end{equation}

\noindent which allows us to see the behaviour as a function of large spacetime separations. Explicitly, the leading IR behaviour of the free Feynman propagator is

\begin{equation}
    \label{free_feynman_prop_leading_IR}
    i\Delta(t,t,\mathbf{0},\mathbf{x})=\frac{H^2}{16\pi^2}\frac{\Gamma(\frac{3}{2}-\nu)\Gamma(2\nu)4^{\frac{3}{2}-\nu}}{\Gamma(\frac{1}{2}+\nu)}\abs{Ha(t)\mathbf{x}}^{-\frac{3}{2}+\nu},
\end{equation}

\noindent where we have focussed on the equal-time propagator. It is this behaviour that one would expect the stochastic approach to reproduce for free fields.

\subsection{Perturbative QFT}

Introducing interactions is a challenging process; the Feynman propagator cannot be found for all values of $m$ and $\lambda$ using current techniques, namely perturbation theory. However, one can still obtain useful results. In perturbative $\phi^4$ theory, the only contribution at  first-order $\lambda$ is the tadpole diagram; therefore, there is no field renormalisation and the only contribution comes via a mass correction, where the bare mass $m$ is replaced by an effective mass $m_{Qeff}$ via

\begin{equation}
    \label{mass_correction}
    m^2\longrightarrow m_{Qeff}^2=m^2+3\lambda\expval{\hat{\phi}^2}.
\end{equation}

\noindent One can therefore obtain the resummed one-loop Feynman propagator by replacing $m$ with $m_{Qeff}$ in Eq. (\ref{qft_feynman_propagator}). The leading IR behaviour is given by

\begin{equation}
    \label{O(lambda)_feynman_prop_leading_IR}
    i\Delta(t,t',\mathbf{x},\mathbf{x}')=\frac{H^2}{16\pi^2}\frac{\Gamma(\frac{3}{2}-\nu_{Qeff})\Gamma(2\nu_{Qeff})}{\Gamma(\frac{1}{2}+\nu_{Qeff})}\qty(-\frac{y}{2})^{-\frac{3}{2}+\nu_{Qeff}},
\end{equation}

\noindent where $\nu_{Qeff}=\sqrt{\frac{9}{4}-\frac{m^2_{Qeff}}{H^2}}$. In practice, this is problematic as $\expval{\hat{\phi}^2}$ contains divergences at $\mathcal{O}(\lambda)$. To see this, we expand Eq. (\ref{qft_feynman_propagator}) for $y\longrightarrow0$ such that the field variance is given by

\begin{equation}
    \label{quantum_field_variance}
    \begin{split}
    \expval{\hat{\phi}^2}=&\lim_{y\rightarrow0}i\Delta(t,t',\mathbf{x},\mathbf{x}')\\=&-\frac{H^2}{8\pi^2y}+\frac{m^2-2H^2}{16\pi^2}\qty(\ln y+i\pi-1+2\gamma_E-\ln2+\psi^{(0)}\qty(\frac{3}{2}-\nu)+\psi^{(0)}\qty(\frac{3}{2}+\nu)),
    \end{split}
\end{equation}

\noindent where $\psi^{(0)}(z)$ are the polygamma functions and $\gamma_E$ is the Euler-Mascheroni constant. We see that the $\mathcal{O}(1/y)$ and $\mathcal{O}(\ln y)$ are ultraviolet (UV) divergent. Such terms are removed by introducing a renormalised mass $m_R^2=m_{0,R}^2+12\xi_RH^2$ such that the divergent $\mathcal{O}(m^2)$ and $\mathcal{O}(H^2)$ are renormalised by the $m_{0,R}$ and $\xi_R$ parameters respectively. Expanding the finite part to leading order in mass, the UV-finite effective mass is given by \cite{garbrecht_rigopoulos:2014}

\begin{equation}
    \label{quantum_effective_mass_light_fields}
    m_{Qeff}^2=m_R^2+\frac{9\lambda H^4}{8\pi^2m_R^2}+\mathcal{O}\qty(\lambda H^2),
\end{equation}

\noindent where the $\mathcal{O}\qty(\lambda H^2)$ part will include terms that are dependent on the renormalisation scheme. This expression tells us that there also exists an infrared (IR) divergence in the theory since the leading term in the small-mass expansion is of relative order $\mathcal{O}\qty(\frac{\lambda H^4}{m^4})$, which is large if $m\ll \lambda^{1/4}H$, and therefore perturbative QFT is only valid in the limit $\lambda\ll m^4/H^4$; the additional finite terms in Eq. (\ref{quantum_effective_mass_light_fields}) are unimportant. Thus, the leading term in the spacelike Feynman propagator is

\begin{equation}
    \label{Feynman_propagator_O(lambda H^4/m^4)}
    i\Delta(t,t,\mathbf{0},\mathbf{x})=\qty(\frac{H^2}{16\pi^2}\frac{\Gamma\qty(\frac{3}{2}-\nu)\Gamma\qty(2\nu)4^{\frac{3}{2}-\nu}}{\Gamma\qty(\frac{1}{2}+\nu)}-\frac{27\lambda H^8}{64\pi^4m_R^6}+\mathcal{O}\qty(\frac{\lambda H^6}{m_R^4}))\abs{Ha(t)\mathbf{x}}^{-\frac{2\Lambda_1^{(QFT)}}{H}},
\end{equation}

\noindent where

\begin{equation}
    \label{QFT_leading eigenvalue}
    \Lambda_1^{(QFT)}=\qty(\frac{3}{2}-\nu)H+\frac{3\lambda H^3}{8\pi^2m_R^2}+\mathcal{O}(\lambda H).
\end{equation}

\noindent It is because of these IR divergences that we pursue other methods of computing correlators in de Sitter, such as the stochastic approach, in order to go beyond the limit where $\lambda\ll m^4/H^4$.

\section{The stochastic approach}
\label{sec:stochastic_approach}

\subsection{The overdamped stochastic approach}

The seminal work of Starobinsky and Yokoyama \cite{starobinsky:1986,Starobinsky-Yokoyama:1994} introduced an effective theory for scalar fields in de Sitter which goes beyond the perturbative methods introduced above. We call this the overdamped (OD) stochastic approach. The principle is that one can separate the short and long wavelength modes of the scalar field such that the short wavelength modes contribute a stochastic noise to the classical equations of motion. Thus, we obtain the IR behaviour of the fields. In the overdamped limit $\Dot{\pi}\ll 3H\pi$ and $V''(\phi)\ll H^2$, one can derive the stochastic equations from the underlying QFT by introducing a strict cutoff between long and short wavelength modes, resulting in the OD stochastic equation

\begin{equation}
    \label{OD_Langevin}
    \Dot{\phi}+\frac{V'(\phi)}{3H}=\xi_{OD}
\end{equation}

\noindent with a white noise contribution $\xi_{OD}(t,\mathbf{x})$ with correlation

\begin{equation}
    \label{OD_noise_corr}
    \expval{\xi_{OD}(t,\mathbf{x})\xi_{OD}(t',\mathbf{x})}=\frac{H^3}{4\pi^2}\delta(t-t').
\end{equation}

\noindent From this starting point, one can do fully non-perturbative calculations to find stochastic correlators that go beyond perturbative QFT. The details of this calculation are given in Ref. \cite{Markkanen:2019} where the authors use a spectral expansion method to perform their computation. To compare this approach with QFT, we write the spacelike OD stochastic field correlator to one-loop order as

\begin{equation}
    \label{OD_field_correlator}
    \expval{\phi(t,\mathbf{0})\phi(t,\mathbf{x})}=\qty(\frac{3H^4}{8\pi^2m^2}-\frac{27\lambda H^8}{64\pi^4m^6}+\mathcal{O}\qty(\lambda^2))\abs{Ha(t)\mathbf{x}}^{-\frac{2\Lambda_1^{(OD)}}{H}},
\end{equation}

\noindent where

\begin{equation}
    \label{OD_eigenvalue}
    \Lambda_1^{(OD)}=\frac{m^2}{3H}+\frac{3\lambda H^3}{8\pi^2m^2}+\mathcal{O}\qty(\lambda^2).
\end{equation}

\noindent Comparing this to the Feynman propagator (\ref{Feynman_propagator_O(lambda H^4/m^4)}), we see that there will only be agreement in the limit where $m\ll H$. The OD stochastic approach doesn't reproduce the full expression for the free part of the Feynman propagator, nor will it include any terms of relative order $\mathcal{O}\qty(\frac{\lambda H^2}{m^2})$ since the next order in the perturbative expansion goes straight to $\mathcal{O}(\lambda^2)$. Thus, the OD stochastic approach is only valid in the regime $\lambda\ll m^2/H^2$.

\subsection{The second-order stochastic equations}
\label{subsec:2nd-order_stochastic_equations}

For the cosmological applications of spectator fields in de Sitter, we wish to go beyond the OD approximation where $m^2\ll H^2$. To do this, we introduce second-order stochastic equations; however, the standard cut-off procedure used to derive the OD stochastic equation is no longer valid when the fields become more massive \cite{Cable:2021}. Instead of attempting a derivation from the underlying QFT, we will introduce a top-down method where we derive stochastic correlation functions from a stochastic equation and then show that these can reproduce their quantum counterparts given the appropriate choice of stochastic parameters. Taking inspiration from Eq. (\ref{scalar_eom}), we write the second-order stochastic equation as

\begin{equation}
    \label{phi_pi_langevin}
    \begin{pmatrix}\Dot{\phi}\\\Dot{\pi}\end{pmatrix}=\begin{pmatrix}\pi\\-3H\pi-V'(\phi)\end{pmatrix}+\begin{pmatrix}\xi_\phi\\\xi_\pi\end{pmatrix},
\end{equation}

\noindent where the potential is given by

\begin{equation}
    \label{stochastic_potential}
    V(\phi)=\frac{1}{2}m^2\phi^2+\frac{1}{4}\lambda \phi^4
\end{equation}

\noindent and the stochastic white noise contributions $\xi_i(t,\mathbf{x})$, $i\in\{\phi,\pi\}$ satisfy

\begin{equation}
    \label{phi_pi_noise_correlation}
    \expval{\xi_i(t,\mathbf{x})\xi_j(t',\mathbf{x})}=\sigma_{ij}^2\delta(t-t').
\end{equation}

\noindent The parameters of the stochastic theory are $m$, $\lambda$ and $\sigma_{ij}^2$. In this paper, we determine their relation to QFT parameters using perturbation theory. Since the perturbative expansion is in powers of $\lambda$, we assume that the couplings in the stochastic approach and perturbative QFT are the same. On the other hand, $m$ and $\sigma_{ij}^2$ will be determined by matching stochastic correlators to their QFT counterparts.

Now that we have a stochastic theory, we introduce the one-point probability distribution function (1PDF) in phase space $P(\phi,\pi;t)$. Its time-evolution is described by the Fokker-Planck equation

\begin{equation}
    \label{phi-pi_fokker-planck_eq}
    \begin{split}
        \partial_t P(\phi,\pi;t)=&\Bigg[3H-\pi\partial_\phi+(3H\pi+V'(\phi))\partial_\pi+\frac{1}{2}\sigma_{\phi\phi}^2\partial_\phi^2+\sigma_{\phi\pi}^2\partial_\phi\partial_\pi+\frac{1}{2}\sigma_{\pi\pi}^2\partial_\pi^2\Bigg]P(\phi,\pi;t)\\
        =&\mathcal{L}_{FP}P(\phi,\pi;t),
    \end{split}
\end{equation}

\noindent where $\mathcal{L}_{FP}$ is called the Fokker-Planck operator.

\subsection{The spectral expansion}
\label{subsec:spectral_expansion}

For free fields, one can solve the Fokker-Planck equation (\ref{phi-pi_fokker-planck_eq}) analytically via a spectral expansion \cite{Cable:2021}. The introduction of self-interactions will result in the need for numerical calculations but one can still use the spectral expansion for the basis of these computations.

We define a space of functions $\{f|(f,f)<\infty\}$ with the scalar product

\begin{equation}
    \label{scalar_product}
    (f,g)=\int_{-\infty}^{\infty} d\phi\int_{-\infty}^{\infty} d\pi f(\phi,\pi)g(\phi,\pi).
\end{equation}

\noindent Note that all integrals over $\phi$ and $\pi$ have the above limits unless otherwise stated. There then exists an adjoint to the Fokker-Planck operator, $\mathcal{L}_{FP}^*$, which is defined via

\begin{equation}
    \label{adjoint_fp_op_definition}
    \qty(\mathcal{L}_{FP}f,g)=\qty(f,\mathcal{L}_{FP}^*g).
\end{equation}

\noindent Explicitly, the adjoint Fokker-Planck operator is

\begin{equation}
    \label{adjoint_fokker-planck_operator}
    \begin{split}
        \mathcal{L}_{FP}^*=&\pi\partial_{\phi}-\qty(3H\pi+V'(\phi))\partial_{\pi}+\frac{1}{2}\sigma_{\phi\phi}^2\partial_{\phi}^2+\sigma_{\phi\pi}^2\partial_{\phi}\partial_{\pi}+\frac{1}{2}\sigma_{\pi\pi}^2\partial_{\pi}^2.
    \end{split}    
\end{equation}

\noindent The 1PDF can be written as

\begin{equation}
    \label{spectral_expansion_1PDF}
    P(\phi,\pi;t)=\Psi_{0}^*(\phi,\pi)\sum_{N=0}^{\infty}\Psi_{N}(\phi,\pi)e^{-\Lambda_{N}t},
\end{equation}

\noindent where $\Lambda_{N}$ and $\Psi_{N}^{(*)}(\phi,\pi)$ are the respective eigenvalues and (adjoint) eigenvectors to the (adjoint) Fokker-Planck operator

\begin{subequations}
    \label{FP_eigenequations}
    \begin{align}
        \mathcal{L}_{FP}\Psi_{N}(\phi,\pi)&=-\Lambda_{N}\Psi_{N}(\phi,\pi),\\
        \mathcal{L}_{FP}^*\Psi_{N}^*(\phi,\pi)&=-\Lambda_{N}\Psi^*_{N}(\phi,\pi).
    \end{align}
\end{subequations}

\noindent The eigenvectors obey the biorthogonality and completeness relations

\begin{subequations}
    \label{biorthogonal&completeness_relations}
    \begin{align}
        \qty(\Psi_{N}^*,\Psi_{N'})&=\delta_{N'N},\\
        \sum_{N}\Psi_{N}^*(\phi,\pi)\Psi_{N}(\phi',\pi')&=\delta(\phi-\phi')\delta(\pi-\pi')
    \end{align}
\end{subequations}

\noindent and there exists an equilibrium state $P_{eq}(\phi,\pi)=\Psi_{0}^*(\phi,\pi)\Psi_{0}(\phi,\pi)$ obeying $\partial_tP_{eq}(\phi,\pi)=0$.

\subsection{Stochastic correlators}

To obtain stochastic correlators, we introduce the transfer matrix $U(\phi_0,\phi,\pi_0,\pi;t)$ between $(\phi_0,\pi_0)=(\phi(0,\mathbf{x}),\pi(0,\mathbf{x}))$ and $(\phi,\pi)=(\phi(t,\mathbf{x}),\pi(t,\mathbf{x}))$, which is defined as the Green's function of the Fokker-Planck equation i.e.

\begin{equation}
    \label{time-evolution_op_FP_equation}
    \partial_tU(\phi_0,\phi,\pi_0,\pi)=\mathcal{L}_{FP}U(\phi_0,\phi,\pi_0,\pi;t)
\end{equation}

\noindent for all values of $\phi_0$ and $\pi_0$. Then, the time-dependence of the 1PDF is given by

\begin{equation}
    \label{time-dependence_1PDF}
    P(\phi,\pi;t)=\int d\phi_0\int d\pi_0 P(\phi_0,\pi_0;0)U(\phi_0,\phi,\pi_0,\pi;t).
\end{equation}

From Eq. (\ref{spectral_expansion_1PDF}), making use of the relations (\ref{biorthogonal&completeness_relations}), we find that the transfer matrix can be written with the spectral expansion as

\begin{equation}
    \label{spectral_expansion_time-ev_operator}
    U(\phi_0,\phi,\pi_0,\pi;t)=\frac{\Psi_{0}^*(\phi,\pi)}{\Psi_{0}^*(\phi_0,\pi_0)}\sum_{N}\Psi_{N}^*(\phi_0,\pi_0)\Psi_{N}(\phi,\pi)e^{-\Lambda_{N}t}
\end{equation}

\noindent and in turn we can write a two-point probability distribution function (2PDF) as

\begin{equation}
    \label{2PDF}
    \begin{split}
        P_2(\phi_0,\phi,\pi_0,\pi;t)&=P_{eq}(\phi_0,\pi_0)U(\phi_0,\phi,\pi_0,\pi;t)\\
        &=\Psi_{0}^*(\phi,\pi)\Psi_{0}(\phi_0,\pi_0)\sum_{N}\Psi_{N}^*(\phi_0,\pi_0)\Psi_{N}(\phi,\pi)e^{-\Lambda_{N}t}.
    \end{split}
\end{equation}

\noindent The 2-point timelike (equal-space) stochastic correlator between some functions $f(\phi_0,\pi_0)$ and $g(\phi,\pi)$ is given by

\begin{equation}
    \label{2pt_timelike_stochastic_correlator_fg}
    \begin{split}
        \expval{f(\phi_0,\pi_0)g(\phi,\pi)}=&\int d\phi_0\int d\phi\int d\pi_0\int d\pi P_2(\phi_0,\phi,\pi_0,\pi;t)f(\phi_0,\pi_0)g(\phi,\pi)\\
        =&\sum_{N}f_{N}^*g_{N}e^{-\Lambda_{N}t},
    \end{split}
\end{equation}

\noindent where
\begin{subequations}
    \label{fnl*&gnl}
    \begin{align}
        f_{N}^*&=\qty(\Psi_{0}f,\Psi_{N}^*),\\
        g_{N}&=\qty(\Psi_{N}g,\Psi_{0}^*).
    \end{align}
\end{subequations}

\noindent In a similar vein to the 2PDF, we can write a three-point probability distribution function (3PDF) between points $(\phi_0,\pi_0)$, $(\phi_1,\pi_1)$ and $(\phi_2,\pi_2)$ as

\begin{equation}
    \label{3PDF_spectral_expansion}
    \begin{split}
        P_3(\phi_0,\phi_1,\phi_2,\pi_0,\pi_1,\pi_2;t_1,t_2)=&P_{eq}(\phi_0,\pi_0)U(\phi_0,\phi_1,\pi_0,\pi_1;t_1)U(\phi_0,\phi_2,\pi_0,\pi_2;t_2)\\
        =&\frac{\Psi_{0}(\phi_0,\pi_0)\Psi^*_{0}(\phi_1,\pi_1)\Psi^*_{0}(\phi_2,\pi_2)}{\Psi^*_{0}(\phi_0,\pi_0)}\sum_{N}\Psi_{N}^*(\phi_0,\pi_0)\Psi_{N}(\phi_1,\pi_1)\\&\times\sum_{N'}\Psi_{N'}^*(\phi_0,\pi_0)\Psi_{N'}(\phi_2,\pi_2)e^{-\qty(\Lambda_{N}t_1+\Lambda_{N'}t_2)}.
    \end{split}
\end{equation}

\noindent To evaluate the spacelike (equal-time) stochastic correlators, we follow Ref. \cite{Starobinsky-Yokoyama:1994} and introduce the time coordinate $t_r$ at which the comoving $\mathbf{x}_1$ and $\mathbf{x}_2$ are inside the same Hubble volume,

\begin{equation}
    \label{space_time_coordinate}
    t_r=-\frac{1}{H}\ln\qty(H\abs{\mathbf{x}_2-\mathbf{x}_1}).
\end{equation}

\noindent Using the 3PDF $P_3(\phi_r,\phi_1,\phi_2,\pi_r,\pi_1,\pi_2)$, the spacelike stochastic correlator between the functions $f(\phi(t,\mathbf{x}_1),\pi(t,\mathbf{x}_1))$ and $g(\phi(t,\mathbf{x}_2),\pi(t,\mathbf{x}_2))$ is given by integrating over $\phi_r$ and $\pi_r$ as

\begin{equation}
    \label{2pt_spacelike_stochastic_correlators_fg}
    \begin{split}
        \langle f&(\phi,\pi;t,\mathbf{x}_1)g(\phi,\pi;t,\mathbf{x}_2)\rangle\\&=\int d\phi_r\int d\phi_1\int d\phi_2\int d\pi_r\int d\pi_1\int d\pi_2 P_3(\phi_r,\phi_1,\phi_2,\pi_r,\pi_1,\pi_2)f(\phi_1,\pi_1)g(\phi_2,\pi_2)
        \\&=\int d\phi_r\int d\pi_r \frac{\Psi_{0}(\phi_r,\pi_r)}{\Psi_{0}^*(\phi_r,\pi_r)}\sum_{NN'}\Psi_{N}^*(\phi_r,\pi_r)\Psi_{N'}^*(\phi_r,\pi_r)f_{N}g_{N'}\abs{Ha(t)(\mathbf{x}_1-\mathbf{x}_2)}^{-\frac{\Lambda_{N}+\Lambda_{N'}}{H}}.
    \end{split}
\end{equation}

\section{Comparison with perturbative QFT}
\label{sec:comparison_with_perturbative_QFT}

\subsection{Free field theory}
\label{subsec:free_fields_comparison}

We now have a formalism that can be used to calculate stochastic correlators but we are yet to attribute it to anything physical as we have not yet specified the noise amplitudes. We now compare it with the QFT results from Sec. \ref{sec:qft}, starting with free fields. This case was studied in full in Ref. \cite{Cable:2021} so this section will be a review.

It will prove convenient to change our field variables from $(\phi,\pi)$ to $(q,p)$, with the transformation

\begin{equation}
    \label{(q,p)=A(phi,pi)}
    \begin{pmatrix}p\\q\end{pmatrix}=\frac{1}{\sqrt{1-\frac{\alpha}{\beta}}}\begin{pmatrix}1&\alpha H\\\frac{1}{\beta H}&1\end{pmatrix}\begin{pmatrix}\pi\\\phi\end{pmatrix},
\end{equation}

\noindent where $\alpha=\frac{3}{2}-\nu$ and $\beta=\frac{3}{2}+\nu$. All of the formalism introduced in the previous section can also be applied to $(q,p)$ variables. In particular the Fokker-Planck operators are given by

\begin{subequations}
    \label{free+int_FP_operators}
    \begin{align}
        \mathcal{L}_{FP}&=\mathcal{L}_{FP}^{(0)}+\lambda\mathcal{L}_{FP}^{(1)},\\
        \mathcal{L}_{FP}^*&=\mathcal{L}_{FP}^{(0)*}+\lambda\mathcal{L}_{FP}^{(1)*},
    \end{align}\\
\end{subequations}

\noindent where the free part is given by

\begin{subequations}
    \label{free_FP_operators}
    \begin{align}
        \mathcal{L}_{FP}^{(0)}&=\alpha H+\alpha H q\partial_q+\frac{1}{2}\sigma_{qq}^{2}\partial_q^2+\beta H +\beta H p\partial_p+\frac{1}{2}\sigma_{pp}^{2}\partial_p^2+\sigma_{qp}^{2}\partial_q\partial_p,\\
        \mathcal{L}_{FP}^{(0)*}&=-\alpha H q\partial_q+\frac{1}{2}\sigma_{qq}^{2}\partial_q^2-\beta H p\partial_p+\frac{1}{2}\sigma_{pp}^{2}\partial_p^2+\sigma_{qp}^{2}\partial_q\partial_p
    \end{align}
\end{subequations}

\noindent and the interacting part is given by

\begin{subequations}
    \label{interacting_FP_operators}
    \begin{align}
        \mathcal{L}_{FP}^{(1)}&=\frac{\lambda}{(1-\frac{\alpha}{\beta})^2}\qty(-\frac{1}{\beta H}p+q)^3\qty(\partial_p+\frac{1}{\beta H}\partial_q),\\
        \mathcal{L}_{FP}^{(1)*}&=-\mathcal{L}_{FP}^{(1)}.
    \end{align}
\end{subequations}

\noindent The $(q,p)$ noise amplitudes are written in terms of their $(\phi,\pi)$ counterparts as

\begin{subequations}
    \label{sigma_q,p-->sigma_phi,pi}
    \begin{align}
        \sigma_{qq}^2=&\frac{1}{1-\frac{\alpha}{\beta}}\qty(\frac{1}{\beta^2H^2}\sigma_{\pi\pi}^2+\frac{2}{\beta H}\sigma_{\phi\pi}^2+\sigma_{\phi\phi}^2),\\
        \sigma_{qp}^2=&\frac{1}{1-\frac{\alpha}{\beta}}\qty(\frac{1}{\beta H}\sigma_{\pi\pi}^2+\qty(1+\frac{\alpha}{\beta })\sigma_{\phi\pi}^2+\alpha H\sigma_{\phi\phi}^2),\\
        \sigma_{pp}^2=&\frac{1}{1-\frac{\alpha}{\beta}}\qty(\sigma_{\pi\pi}^2+2\alpha H\sigma_{\phi\pi}^2+\alpha^2H^2\sigma_{\phi\phi}^2).
    \end{align}
\end{subequations}

\noindent Further, the $\phi-\phi$ correlator is written in terms of the $(q,p)$ correlators as

\begin{small}
\begin{equation}
    \label{phi-phi_pq}
    \begin{split}
    \expval{\phi(t,\mathbf{x})\phi(t',\mathbf{x}')}=\frac{1}{1-\frac{\alpha}{\beta}}\Bigg(&\frac{1}{\beta^2H^2}\expval{p(t,\mathbf{x})p(t',\mathbf{x}')}-\frac{1}{\beta H}\expval{q(t,\mathbf{x})p(t',\mathbf{x}')}\\&-\frac{1}{\beta H}\expval{p(t,\mathbf{x})q(t',\mathbf{x}')}+\expval{q(t,\mathbf{x})q(t',\mathbf{x}')}\Bigg).
    \end{split}
\end{equation}
\end{small}

\noindent Similar expressions can be found for the $\phi-\pi$, $\pi-\phi$ and $\pi-\pi$ correlators but we will focus on the $\phi-\phi$ correlator in this work. Following the work of Ref. \cite{Cable:2021}, we compute the stochastic free field correlator and match it to the free Feynman propagator (\ref{free_feynman_prop_leading_IR}) to obtain an expression for the noise amplitudes, resulting in

\begin{subequations}
    \label{matched_(p,q)_noise_free}
\begin{align}
        \label{matched_sigma_qq_free}
        \sigma_{qq}^{2(0)}&=\frac{2H^3\Gamma(1+\nu)\Gamma\qty(\frac{5}{2}-\nu)}{\pi^{5/2}\qty(3+2\nu)},\\
        \label{matched_sigma_qp_free}
        \sigma_{qp}^{2(0)}&=0,\\
        \label{matched_sigma_pp_free}
        \sigma_{pp}^{2(0)}&=0.
\end{align}
\end{subequations}

The $qq$ noise is matched such that the leading-order term in the Feynman propagator is reproduced and the $qp$ noise is chosen such that there is an analytic continuation from timelike to spacelike stochastic correlators, a behaviour prevalent in QFT. However, the choice of $\sigma_{pp}^2$ is arbitrary. In this paper, we set it to be zero - the subleading term doesn't contribute - because it is the simplest case. The $(q,p)$ noise matrix contains a zero eigenvalue and thus one would expect the stochastic equations to simplify, though we don't pursue this here.

In Ref. \cite{Cable:2021} we use a different choice, where the subleading contribution reproduces the leading term in the second sum $\propto \abs{H a(t)\mathbf{x}}^{-\frac{3}{2}-\nu}$. We denote this by $\sigma_{pp}^{2(NLO)}$ and will be used for comparison with Eq. (\ref{matched_sigma_pp_free}) in Sec. \ref{sec:numerical_calculation}. We will discuss any differences this makes to the conclusions drawn in Ref. \cite{Cable:2021} in Appendix \ref{app:pp_noise_amplitude}. Henceforth, we will use Eq. (\ref{matched_(p,q)_noise_free}) as our free noise but we include a more general formalism for calculating stochastic field correlators in Appendix \ref{app:perturbation_theory_general_noise}.

Since $\sigma_{qp}^{2(0)}=0$, the variables $p$ and $q$ separate and so we now use two indices $(r,s)\in\{0,\infty\}$, corresponding to $p$ and $q$ respectively, as opposed to just $N$. Thus, the free field eigenquations are given by

\begin{subequations}
    \label{FP_free_eigenequations}
    \begin{align}
        \mathcal{L}^{(0)}_{FP}\Psi_{rs}^{(0)}(q,p)&=-\Lambda_{rs}^{(0)}\Psi_{rs}^{(0)}(q,p),\\
        \mathcal{L}_{FP}^{(0)*}\Psi_{rs}^{(0)*}(q,p)&=-\Lambda_{rs}^{(0)}\Psi^{(0)*}_{rs}(q,p),
    \end{align}
\end{subequations}

\noindent where the $\Lambda_{rs}^{(0)}$ and $\Psi_{rs}^{(0)(*)}(q,p)$ are the free eigenvalues and (adjoint) eigenstates respectively. The eigenvalues of Eq. (\ref{FP_free_eigenequations}) are

\begin{equation}
    \label{free_eigenvalues}
    \Lambda_{rs}^{(0)}=\qty(s\alpha+r\beta)H
\end{equation}

\noindent while the normalised eigenstates can be written in terms of the Hermite polynomials $H_n(x)$ as

\begin{subequations}
    \label{free_eigenstates}
    \begin{align}
        \label{free_non-adjoint_estate}
        \Psi_{rs}^{(0)}(q,p)&=\frac{1}{\sqrt{2^{r+s}r!s!}}\qty(\frac{\alpha\beta H^2}{\pi^2\sigma_{qq}^{2}\sigma_{pp}^{2}})^{1/4}H_s\qty(\sqrt{\frac{\alpha H}{\sigma_{qq}^{2}}}q)H_r\qty(\sqrt{\frac{\beta H}{\sigma_{pp}^{2}}}p)e^{-\frac{\alpha H}{\sigma_{qq}^{2}}q^2-\frac{\beta H}{\sigma_{pp}^{2}}p^2},\\
        \label{free_adjoint_estate}
        \Psi_{rs}^{(0)*}(q,p)&=\frac{1}{\sqrt{2^{r+s}r!s!}}\qty(\frac{\alpha\beta H^2}{\pi^2\sigma_{qq}^{2}\sigma_{pp}^{2}})^{1/4}H_s\qty(\sqrt{\frac{\alpha H}{\sigma_{qq}^{2}}}q)H_r\qty(\sqrt{\frac{\beta H}{\sigma_{pp}^{2}}}p).
    \end{align}
\end{subequations}

\noindent For the case where $\sigma_{pp}^2=0$, the eigenstates can be written as\footnote{To take the limit, we have used the identity $\lim_{\epsilon\rightarrow0}\frac{(-1)^{-n}(\sqrt{2}\epsilon)^{n-1}}{\sqrt{\pi}}H_n\qty(\frac{x}{\sqrt{2}\epsilon})e^{-\frac{x^2}{2\epsilon^2}}=\delta^{(n)}(x)$.}

\begin{subequations}
    \label{free_eigenstates_spp=0}
    \begin{align}
        \label{free_non-adjoint_estate_spp=0}
        \lim_{\sigma_{pp}^2\rightarrow0}\Psi_{rs}^{(0)}(q,\Tilde{p})&=\frac{(-1)^{-r}}{\sqrt{2^{r+s}r!s!}}\qty(\frac{\alpha H}{\sigma_{qq}^2})^{1/4}\delta^{(r)}(\Tilde{p})H_s\qty(\sqrt{\frac{\alpha H}{\sigma_{qq}^2}q})e^{-\frac{\alpha H}{\sigma_{qq}^2}q^2},\\
        \label{free_adjoint_estate_spp=0}
        \lim_{\sigma_{pp}^2\rightarrow0}\Psi_{rs}^{(0)*}(q,\Tilde{p})&=\sqrt{\frac{2^r}{2^sr!s!}}\qty(\frac{\alpha H}{\pi^2\sigma_{qq}^2})^{1/4}\Tilde{p}^rH_s\qty(\sqrt{\frac{\alpha H}{\sigma_{qq}^2}q}),
    \end{align}\\
\end{subequations}

\noindent where $\Tilde{p}=\sqrt{\frac{\beta H}{\sigma_{pp}^2}}p$ and superscript $(r)$ indicates we are taking the $r$th derivative of the $\delta$-function. These are well behaved eigenstates if we use $(q,\Tilde{p})$ as our variables, with which we have the biorthogonality and completeness relations.

\subsection{Stochastic perturbation theory}
\label{subsec:stochastic_perturbation_theory}

We will now move to the more interesting case of an interacting theory. To relate the stochastic correlators to the perturbative results of QFT, we expand our solutions to the eigenproblem (\ref{FP_eigenequations}) in terms of the $(q,p)$ variables to $\mathcal{O}(\lambda)$

\begin{subequations}
    \label{perturbed_eigensolutions}
    \begin{align}
        \Lambda_{rs}=&\Lambda_{rs}^{(0)}+\lambda \Lambda_{rs}^{(1)},\\
        \Psi_{rs}^{(*)}(q,p)=&\Psi_{rs}^{(0)(*)}(q,p)+\lambda\Psi_{rs}^{(1)(*)}(q,p).
    \end{align}
\end{subequations}  

\noindent Using the eigenequations with the biorthogonality conditions for $(q,p)$, equivalent to  Eq. (\ref{FP_eigenequations}) and (\ref{biorthogonal&completeness_relations}), the $\mathcal{O}(\lambda)$ terms in the eigenvalues and eigenstates are given by

\begin{subequations}
    \label{O(lambda)_eigenvalues&eigenstates}
    \begin{align}
    \label{O(lambda)_eigenvalues}
        \Lambda_{rs}^{(1)}&=-\qty(\Psi_{rs}^{(0)*},\mathcal{L}_{FP}^{(1)}\Psi_{rs}^{(0)}),\\
    \label{O(lambda)_eigenstates}  
        \Psi_{rs}^{(1)}(q,p)&=\sum_{r's'}\Psi_{r's'}^{(0)}(q,p)\frac{\qty(\Psi_{r's'}^{(0)*},\mathcal{L}_{FP}^{(1)}\Psi_{rs}^{(0)})}{\Lambda_{r's'}^{(0)}-\Lambda_{rs}^{(0)}},\\ 
    \label{O(lambda)_adjoint_eigenstates}   
        \Psi_{rs}^{(1)*}(q,p)&=\sum_{r's'}\Psi_{r's'}^{(0)*}(q,p)\frac{\qty(\Psi_{r's'}^{(0)},\mathcal{L}_{FP}^{(1)*}\Psi_{rs}^{(0)*})}{\Lambda_{r's'}^{(0)}-\Lambda_{rs}^{(0)}}, 
    \end{align}
\end{subequations}

\noindent where for Eq. (\ref{O(lambda)_eigenstates}) and (\ref{O(lambda)_adjoint_eigenstates}), $r'\ne r$ and $s'\ne s$. 
By applying the expansion (\ref{perturbed_eigensolutions}) to Eq. (\ref{2pt_timelike_stochastic_correlator_fg}), we can write the timelike correlator between two functions $f$ and $g$ to $\mathcal{O}(\lambda)$ as

\begin{equation}
    \label{O(lambda)_timelike_stochastic_correlator}
    \expval{f(q_0,p_0)g(q,p)}=\sum_{rs}\qty[f^{(0)*}_{rs}g_{rs}^{(0)}+\lambda\qty(f_{rs}^{(0)*}g_{rs}^{(1)}+f_{rs}^{(1)*}g_{rs}^{(0)})]e^{-\qty(\Lambda_{rs}^{(0)}+\lambda\Lambda_{rs}^{(1)})t}
\end{equation}

\noindent where $g_{rs}=g_{rs}^{(0)}+\lambda g_{rs}^{(1)}$ such that

\begin{subequations}
    \begin{align}
        \label{gnl_perturbative_expansion}
        g_{rs}^{(0)}&=\qty(\Psi_{rs}^{(0)}g,\Psi_{00}^{(0)*}),\\
        g_{rs}^{(1)}&=\qty(\Psi_{rs}^{(1)}g,\Psi_{00}^{(0)*})+\qty(\Psi_{rs}^{(0)}g,\Psi_{00}^{(1)*})
    \end{align}
\end{subequations}

\noindent with a similar relation holding for $f_{rs}^*$. A similar expression for the spacelike correlator can be written using Eq. (\ref{2pt_spacelike_stochastic_correlators_fg}) as

\begin{equation}
\label{O(lambda)_spacelike_stochastic_correlator}
    \begin{split}
        \expval{f(q_1,p_1)g(q_2,p_2)}=&\int dq_r\int dp_r\sum_{r'rs's}\Bigg[\frac{\Psi_{00}^{(0)}(q_r,p_r)}{\Psi_{00}^{(0)*}(q_r,p_r)}\Psi_{rs}^{(0)*}(q_r,p_r)\Psi_{r's'}^{(0)*}(q_r,p_r)f_{rs}^{(0)}g_{r's'}^{(0)}\\&
        \begin{split}
        +\lambda\Bigg(&\frac{\Psi_{00}^{(1)}(q_r,p_r)}{\Psi_{00}^{(0)*}(q_r,p_r)}\Psi_{rs}^{(0)*}(q_r,p_r)\Psi_{r's'}^{(0)*}(q_r,p_r)f_{rs}^{(0)}g_{r's'}^{(0)}
        \\&-\frac{\Psi_{00}^{(1)*}(q_r,p_r)\Psi_{00}^{(0)}(q_r,p_r)}{\Psi_{00}^{(0)*}(q_r,p_r)^2}\Psi_{rs}^{(0)*}(q_r,p_r)\Psi_{r's'}^{(0)*}(q_r,p_r)f_{rs}^{(0)}g_{r's'}^{(0)}
        \\&+\frac{\Psi_{00}^{(0)}(q_r,p_r)}{\Psi_{00}^{(0)*}(q_r,p_r)}\Psi_{rs}^{(1)*}(q_r,p_r)\Psi_{r's'}^{(0)*}(q_r,p_r)f_{rs}^{(0)}g_{r's'}^{(0)}
        \\&+\frac{\Psi_{00}^{(0)}(q_r,p_r)}{\Psi_{00}^{(0)*}(q_r,p_r)}\Psi_{rs}^{(0)*}(q_r,p_r)\Psi_{r's'}^{(1)*}(q_r,p_r)f_{rs}^{(0)}g_{r's'}^{(0)}
        \\&+\frac{\Psi_{00}^{(0)}(q_r,p_r)}{\Psi_{00}^{(0)*}(q_r,p_r)}\Psi_{rs}^{(0)*}(q_r,p_r)\Psi_{r's'}^{(0)*}(q_r,p_r)f_{rs}^{(1)}g_{r's'}^{(0)}
        \\&+\frac{\Psi_{00}^{(0)}(q_r,p_r)}{\Psi_{00}^{(0)*}(q_r,p_r)}\Psi_{rs}^{(0)*}(q_r,p_r)\Psi_{r's'}^{(0)*}(q_r,p_r)f_{rs}^{(0)}g_{r's'}^{(1)}\Bigg)\Bigg]
        \end{split}
        \\&\times \abs{Ha(t)(\mathbf{x}_1-\mathbf{x}_2)}^{-\qty(\frac{\Lambda_{rs}^{(0)}+\Lambda_{r's'}^{(0)}}{H}+\lambda\frac{\Lambda_{rs}^{(1)}+\Lambda_{r's'}^{(1)}}{H})}.
    \end{split}
\end{equation}

\noindent It is the spacelike correlator that we will focus on.

We are now in a position where we can substitute explicit expressions for the eigenvalues and eigenstates in terms of the noise amplitudes (\ref{matched_(p,q)_noise_free}). We will only consider the zeroth and first two non-zero states in the spectral expansion since these are the comparable terms with the QFT results. Hence, we only need to concern ourselves with finding the terms in the correlators corresponding to $(r,s)$ equal to $(0,0)$, $(0,1)$ and $(1,0)$. From Eq. (\ref{O(lambda)_eigenvalues&eigenstates}), the $\mathcal{O}(\lambda)$ corrections to the first two eigenvalues are

\begin{subequations}
    \label{lowest_3_O(lambda)_eigenvalues}
    \begin{align}
        \Lambda_{00}^{(1)}&=0,\\
         \Lambda_{01}^{(1)}&=\frac{3H\Gamma(\nu)\Gamma\qty(\frac{3}{2}-\nu)}{8\pi^{5/2}\nu},
    \end{align}
\end{subequations}

\noindent where the fact that $\Lambda_{00}^{(1)}=0$ is consistent with its correspondence to the equilibrium solution. Using Eq. (\ref{O(lambda)_spacelike_stochastic_correlator}), we can find the spacelike $q-q$, $q-p$, $p-q$ and $p-p$ stochastic correlators. To complete our perturbative expansion, we also need to account for the $\lambda$-dependence in the noise by

\begin{equation}
    \label{noise_amplitude_perturbative_expansion}
    \sigma_{ij}^2=\sigma_{ij}^{2(0)}+\lambda\sigma_{ij}^{2(1)}.
\end{equation}

\noindent Note that we will show to relative order $\mathcal{O}\qty(\frac{\lambda H^4}{m^4})$ that $\sigma_{ij}^{2(1)}$ is zero; however, it's important to include at this stage to show that no such IR divergent piece contributes. To $\mathcal{O}(\lambda)$, the spacelike $(q,p)$ stochastic correlators are

\begin{subequations}
    \label{(q,p)_spacelike_stochastic_correlators}
    \begin{align}
    \label{spacelike_qq}
            \expval{q(t,\mathbf{0})q(t,\mathbf{x})}=&\qty[\frac{H^4\Gamma\qty(1+\nu)\Gamma\qty(\frac{5}{2}-\nu)}{2\pi^{5/2}m^2}+\lambda\qty(\frac{\sigma_{qq}^{2(1)}}{H\qty(3-2\nu)}-\frac{3H^4\Gamma(\nu)^2\Gamma\qty(\frac{3}{2}-\nu)^2}{8\pi^5(3+2\nu)m^2})]\abs{Ha(t)\mathbf{x}}^{-\frac{2\Lambda_{01}}{H}},\\
    \label{spacelike_qp}
        \begin{split}
            \expval{q(t,\mathbf{0})p(t,\mathbf{x})}=&\expval{p(t,\mathbf{0})q(t,\mathbf{x})}=-\frac{3\lambda H^3\Gamma(\nu)^2\Gamma\qty(\frac{3}{2}-\nu)^2}{32\pi^5\nu}\abs{Ha(t)\mathbf{x}}^{-\frac{2\Lambda_{01}}{H}}\\&
            \hspace{32mm}+\lambda\qty(\frac{\sigma_{qp}^{2(1)}}{3H}+\frac{H^5\Gamma(\nu)^2\Gamma\qty(\frac{5}{2}-\nu)^2}{16\pi^5\nu m^2})\abs{Ha(t)\mathbf{x}}^{-3},
        \end{split}\\
    \label{spacelike_pp}
        \begin{split}
            \expval{p(t,\mathbf{0})p(t,\mathbf{x})}=&\frac{\lambda \sigma_{pp}^{2(1)}}{H(3+2\nu)}\abs{Ha(t)\mathbf{x}}^{-\frac{2\Lambda_{10}}{H}}.
        \end{split}
    \end{align}
\end{subequations}

\noindent Substituting these expressions into Eq. (\ref{phi-phi_pq}), we obtain an expression for the $\phi-\phi$ stochastic correlator up to first order in $\lambda$. The spacelike version is

\begin{equation}
    \label{phi-phi_spacelike_stochastic_correlator}
    \begin{split}
        \langle\phi(t&,\mathbf{0})\phi(t,\mathbf{x})\rangle\\=&\Bigg[\frac{H^2}{16\pi^2}\frac{\Gamma\qty(\frac{3}{2}-\nu)\Gamma\qty(2\nu)4^{\frac{3}{2}-\nu}}{\Gamma\qty(\frac{1}{2}+\nu)}+\lambda\qty(\frac{(3+2\nu)\sigma_{qq}^{2(1)}}{4\nu H(3-2\nu)}+\frac{3(3-4\nu)H^4\Gamma(\nu)^2\Gamma\qty(\frac{3}{2}-\nu)^2}{32\pi^5\nu m^2})\Bigg]\\&\times\abs{Ha(t)\mathbf{x}}^{-\frac{2\Lambda_{01}}{H}}
        \\&+\frac{\lambda \sigma_{pp}^{2(1)}}{H^3\nu(3+2\nu)^2}\abs{Ha(t)\mathbf{x}}^{-\frac{2\Lambda_{10}}{H}}-
        \lambda\qty(\frac{\sigma_{qp}^{2(1)}}{3H^2\nu}+\frac{H^4\Gamma(\nu)^2\Gamma\qty(\frac{5}{2}-\nu)^2}{8\pi^5\nu m^2})\abs{Ha(t)\mathbf{x}}^{-3}.
    \end{split}
\end{equation}

\subsection{The second-order stochastic parameters}
\label{subsec:O(lambda)_noise}

In Sec. \ref{sec:qft}, we found the Feynman propagator to $\mathcal{O}\qty(\lambda)$. We will use this result to match the $\mathcal{O}(\lambda)$ noise. First, we need to match the stochastic mass $m$ to the parameters $m_R$ and $\lambda$ from the QFT such that the exponents of Eq. (\ref{Feynman_propagator_O(lambda H^4/m^4)}) and (\ref{phi-phi_spacelike_stochastic_correlator}) agree to relative order $\mathcal{O}\qty(\frac{\lambda H^4}{m^4})$. Combining the free and interacting parts of the first non-zero eigenvalue, we find that to relative order $\mathcal{O}\qty(\frac{\lambda H^4}{m^4})$

\begin{equation}
    \label{first_excited_evalue}
    \Lambda_{01}=\qty(\frac{3}{2}-\nu)H+\frac{3\lambda H^3}{8\pi^2m^2}+\mathcal{O}(\lambda H).
\end{equation}

\noindent has the same function form as the exponent in the Feynman propagator (\ref{QFT_leading eigenvalue}). This tells us that the stochastic mass $m^2=m^2_R\qty(1+\mathcal{O}\qty(\frac{\lambda H^2}{m_R^2}))$. In particular, they agree at relative order $\mathcal{O}\qty(\frac{\lambda H^4}{m^4})$ and thus the stochastic exponent can be found for $\lambda\ll m^2/H^2$ in contrast to the direct perturbative calculation which requires $\lambda\ll m^4/H^4$. In order to compute the term at relative order $\mathcal{O}\qty(\frac{\lambda H^2}{m^2_R})$, we would have to choose a regularisation scheme. This will be considered in future work \cite{in_preparation}. For the rest of this paper, we will just set any $\mathcal{O}\qty(\frac{\lambda H^2}{m^2})$ corrections - be they quantum or stochastic - to zero. It is noteworthy that Eq. (\ref{first_excited_evalue}) is also obtained if we were to choose $\sigma_{pp}^2=\sigma_{pp}^{2(NLO)}$.

To match the amplitude, we expand the spacelike stochastic field correlator (\ref{phi-phi_spacelike_stochastic_correlator}) to relative order $\mathcal{O}\qty(\frac{\lambda H^4}{m^4})$ resulting in

\begin{equation}
    \label{phi-phi_spacelike_stochastic_correlator_O(lambdaH4/m4)}
    \begin{split}
    \expval{\phi(t,\mathbf{0})\phi(t,\mathbf{x})}=&\qty[\frac{3H^4}{8\pi m^2}-\frac{27\lambda H^8}{64\pi^4m^6}+\frac{3H\sigma_{qq}^{2(1)}}{2m^2}+\mathcal{O}\qty(\frac{\lambda H^6}{m^4})]\abs{Ha(t)\mathbf{x}}^{-\frac{2m^2}{3H^2}+\frac{3\lambda H^2}{4\pi^2m^2}+\mathcal{O}(\lambda)}
    \\&+\qty(\frac{\lambda \sigma_{pp}^{2(1)}}{54H^3}+\mathcal{O}\qty(\frac{\lambda H^4}{m^2}))\abs{Ha(t)\mathbf{x}}^{-3-\frac{2m^2}{3H^2}-\frac{3\lambda H^2}{4\pi^2m^2}+\mathcal{O}(\lambda)}
    \\&+\lambda\qty(\frac{2\sigma_{qp}^{2(1)}}{9H^2}+\mathcal{O}\qty(\frac{\lambda H^4}{m^2}))\abs{Ha(t)\mathbf{x}}^{-3}.
    \end{split}
\end{equation}

\noindent We see that if we set all three $\mathcal{O}(\lambda)$ noise amplitudes to zero, we will reproduce the Feynman propagator to relative order $\mathcal{O}\qty(\frac{\lambda H^4}{m^4})$. Thus, our matched noise from perturbation theory is

\begin{subequations}
    \label{matched_(p,q)_noise}
\begin{align}
        \label{matched_sigma_qq}
        \sigma_{qq}^{2}&=\frac{2H^3\Gamma(1+\nu)\Gamma\qty(\frac{5}{2}-\nu)}{\pi^{5/2}\qty(3+2\nu)}+\mathcal{O}\qty(\frac{\lambda H^5}{m^2}),\\
        \label{matched_sigma_qp}
        \sigma_{qp}^{2}&=0+\mathcal{O}\qty(\frac{\lambda H^6}{m^2}),\\
        \label{matched_sigma_pp}
        \sigma_{pp}^{2}&=0+\mathcal{O}\qty(\frac{\lambda H^7}{m^2}).
\end{align}
\end{subequations}

\noindent We are assuming that $\sigma_{qp}^{2(0)}$ and $\sigma_{pp}^{2(0)}$ are parametrically $\mathcal{O}\qty(H^4)$ and $\mathcal{O}\qty(H^5)$ respectively such that the correction to all three noise amplitudes is of the same relative order $\mathcal{O}\qty(\frac{\lambda H^2}{m^2})$. Converting the noise amplitudes to $(\phi,\pi)$ variables gives

\begin{equation}
    \label{(phi,pi)_noise_sigmapp=0}
    \sigma^2=\frac{H^3\Gamma\qty(\nu)\Gamma\qty(\frac{5}{2}-\nu)}{2\pi^{5/2}}\begin{pmatrix}
    1+\mathcal{O}\qty(\frac{\lambda H^5}{m^2})&-\frac{2m^2}{H(3+2\nu)}+\mathcal{O}\qty(\frac{\lambda H^6}{m^2})\\-\frac{2m^2}{H(3+2\nu)}+\mathcal{O}\qty(\frac{\lambda H^6}{m^2})&\frac{4m^4}{(3+2\nu)^2H^2}+\mathcal{O}\qty(\frac{\lambda H^7}{m^2})
    \end{pmatrix}.
\end{equation}

\noindent This matrix gives the components of the noise aligning with Eq. (\ref{phi_pi_noise_correlation}).

\subsection{Comparing models and their limitations}

The strength of a stochastic approach is that the Fokker-Planck can be solved - analytically for some examples but mostly numerically - and therefore correlation functions can be obtained non-perturbatively. The second-order stochastic approach loses some of this power because it requires the noise to be calculated perturbatively and therefore we are at present limited to the regime $\frac{\lambda H^2}{m^2}\ll 1$. To investigate the usefulness of this approach, we compare its regime of validity with that of the other two approximations introduced above: perturbative QFT and stochastic OD. Their respective regimes of validity are

\begin{subequations}
    \label{model_limitations}
    \begin{align}
        \text{Perturbative QFT: }\qquad\lambda\ll &\frac{m^4}{H^4}, \qquad \lambda\ll1\\
        \text{OD stochastic: }\qquad\lambda\ll&\frac{m^2}{H^2}, \qquad m\ll H, \\
        \text{Second-order stochastic: }\qquad\lambda\ll &\frac{m^2}{H^2}, \qquad m\lesssim H.
    \end{align}
\end{subequations}

\begin{figure}[ht]
    \centering
    \includegraphics[width=140mm]{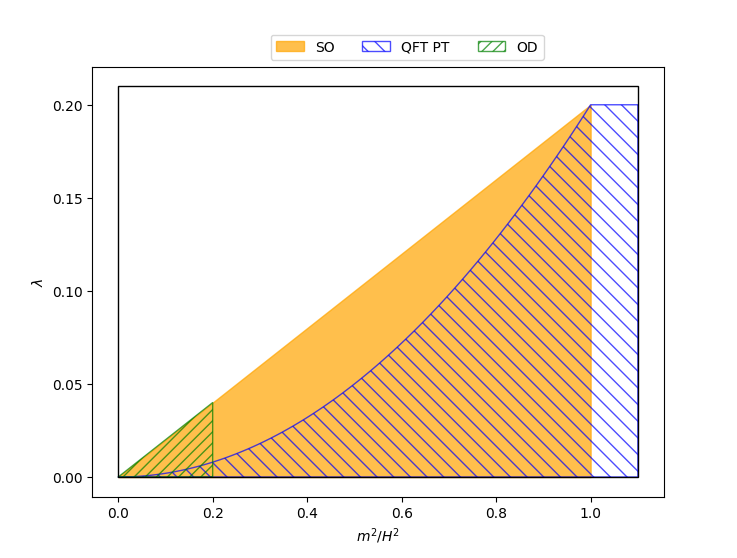}
    \caption{This shows the regimes in which we expect our approximations to work. Perturbative QFT, OD stochastic and second-order (SO) stochastic are expected to work in the blue left hashed, green right hashed and orange regions respectively. Note that there is some overlap. The pure white space is where none of these approximations work.}
    \label{fig:model_limitations_comparison}
\end{figure}

\noindent A comparison between these regimes is given in Fig. \ref{fig:model_limitations_comparison}. For the purposes of making the boundaries obvious, we choose ``$\ll 1$'' to mean ``$<0.2$'' in this plot. In reality, we wouldn't expect these boundaries to be so clear cut.

Perturbative QFT is encompassed by the second-order stochastic approach in the limit $m\lesssim H$, which is unsurprising given that the matched correlators were found directly from the Feynman propagators. The OD results should also be completely covered by the second-order stochastic approach - we will confirm this in Sec. \ref{subsubsec:OD_v_matching_comparison} - and are resigned to the left hand side of Fig. \ref{fig:model_limitations_comparison} as the approximation requires the fields to be light. Importantly, this leaves an area in the parameter space - the purely orange zone - where we expect the second-order stochastic approach to work when the others do not. To obtain such results, we need to numerically solve our 2-d Fokker Planck equation. From this, we can compare the three approximations. 

\section{Numerical solutions to the Fokker-Planck equation}
\label{sec:numerical_calculation}

\subsection{Expansion in free eigenstates}
\label{subsec:expansion_in_free_eigenstates}

Thus far, we have shown that the second-order stochastic approach can be made to coincide with perturbative QFT if we choose the stochastic parameters to be Eq. (\ref{matched_(p,q)_noise}). We will now solve the stochastic equations numerically to obtain non-perturbative results. We continue to use the $(q,p)$ coordinates so that we can continue to use the free eigenstates (\ref{free_eigenstates}). Making the ansatz that our eigensolutions to the eigenequations (\ref{FP_eigenequations}) can be written as

\begin{subequations}
    \label{expansion_free_eigensolutions}
    \begin{align}
        \Psi_{N}(q,p)&=\sum_{rs}c^{(N)}_{rs}\psi_{rs}^{(0)}(q,p),\\
        \Psi_{N}^*(q,p)&=\sum_{rs}c^{*(N)}_{rs}\psi_{rs}^{(0)*}(q,p),
    \end{align}
\end{subequations}

\noindent where $c_{rs}^{(*)(N)}$ are two sets of coefficients to be determined and $\psi_{rs}^{(0)*}(q,p)$ are the free eigenstates given in Eq. (\ref{free_eigenstates}). Substituting Eq. (\ref{expansion_free_eigensolutions}) into (\ref{FP_eigenequations}) gives

\begin{subequations}
    \label{expansion_expression_FP_eigenequation}
    \begin{align}
        \sum_{rs}c_{rs}^{(N)}\mathcal{L}_{FP}\psi_{rs}^{(0)}(q,p)&=-\Lambda_N\sum_{rs}c_{rs}^{(N)}\psi_{rs}^{(0)}(q,p),\\
        \sum_{rs}c_{rs}^{(N)}\mathcal{L}^*_{FP}\psi_{rs}^{(0)*}(q,p)&=-\Lambda_N\sum_{rs}c_{rs}^{*(N)}\psi_{rs}^{(0)*}(q,p).
    \end{align}
\end{subequations}

\noindent Applying the Fokker-Planck operator to the free eigenstates will give us

\begin{equation}
    \label{FP_operator-->matrix}
    \mathcal{L}_{FP}^{(*)}\psi_{rs}^{(0)(*)}(q,p)=\sum_{r's'}\mathcal{M}^{(*)}_{rsr's'}\psi_{r's'}^{(0)(*)},
\end{equation}

\noindent where the matrices $\mathcal{M}^{(*)}$ are given by

\begin{equation}
    \label{matrices_M}
    \mathcal{M}_{rsr's'}=\qty(\psi_{r's'}^{(0)*},\mathcal{L}_{FP}\psi_{rs}^{(0)})=\mathcal{M}_{r's'rs}^*.
\end{equation}

\noindent Explicit expressions for these matrices can be found but they are complicated. Applying Eq. (\ref{FP_operator-->matrix}) to (\ref{expansion_expression_FP_eigenequation}) and making use of the completeness of the free eigenstates (\ref{biorthogonal&completeness_relations}), one can write

\begin{subequations}
    \label{matrix_eigenequation}
    \begin{align}
        \sum_{r's'}\mathcal{M}^T_{rsr's'}c_{r's'}^{(N)}&=-\Lambda_{N}c_{rs}^{(N)},\\
        \sum_{r's'}(\mathcal{M}^*)^T_{rsr's'}c_{r's'}^{*(N)}&=-\Lambda_{N}c_{rs}^{*(N)}.
    \end{align}
\end{subequations}

\noindent Thus, by diagonalising the matrices $(\mathcal{M}^{(*)})^T$, we can obtain the eigenvalues $\Lambda_N$ and the coefficients $c_{rs}^{*(N)}$ and hence the full solution to the Fokker-Planck equation.

In theory, this sum is infinite and our matrices are infinite-dimensional. Therefore, we have to choose a value of $r$ and $s$ ($r_{max}$ and $s_{max}$ respectively) at which we truncate the series so that we can practically diagonalise the matrices. This approximation only works if the expansion in our chosen eigenstates (\ref{expansion_free_eigensolutions}) converges as $r_{max}$ and $s_{max}$ become large. Indeed, we can use this fact to improve the accuracy of the spectral expansion by evaluating the eigenvalues for a range of $r$ and $s$ and then fitting an appropriate curve that converges at infinity. This essentially gives us the eigenvalue at infinity. There will naturally be some error associated with this fit but, as we will see, it is exceedingly small. The convergence speeds up as $m^2/H^2$ increases with constant $\lambda$ and as $\lambda$ decreases with constant $m^2/H^2$. This is the case where the free solution is the dominant one.

\begin{figure}[ht]
    \centering
    \includegraphics[width=150mm]{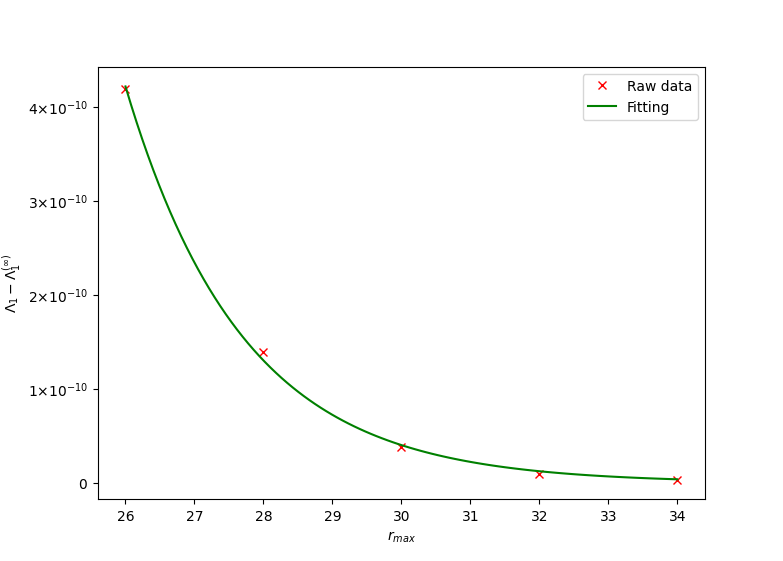}
    \caption{The difference between the first non-zero eigenvalue as found at the highest order of truncation $\Lambda_1$ and the fit $\Lambda_1^{(\infty)}=0.004530997449(4)$ for a range of $(r_{max},s_{max})$ at $m^2/H^2=0.01$ and $\lambda=0.0005$. Red crosses are the data found by numerically diagonalising the matrix $\mathcal{M}^T$ up to the truncation $(r_{max},s_{max})$ and the green line is the fit, which is exponential. Note that we always take $r_{max}=s_{max}$ and hence use a single number, $r_{max}$, to label the $x$-axis.}
    \label{fig:fit_evals_v_trunc}
\end{figure}

For the purposes of this work, we will focus on calculating the first non-zero eigenvalue in our spectral expansion $\Lambda_{1}$. To make the idea of truncation and fitting more concrete, we consider the specific example in Fig. \ref{fig:fit_evals_v_trunc} where we calculate $\Lambda_{1}$ at $m^2/H^2=0.01$ and $\lambda=0.0005$ with the level of truncation $(r_{max},s_{max})$ ranging from (26,26) to (34,34). The fit in Fig. \ref{fig:fit_evals_v_trunc} gives the eigenvalue at infinity $\Lambda_1^{(\infty)}=0.004530997449(4)$. The error is exceedingly small, of order $10^{-12}$. Even if one just studies Fig. \ref{fig:fit_evals_v_trunc} roughly, one can see that the value of $\Lambda_1$ changes on the scale of $10^{-10}$ when going from a truncation at (26,26) to (34,34), 7 orders of magnitude below the leading significant figure of the eigenvalue. This is so small that we can consider our numerical approach to have negligible error. This is the scale of errors for all data taken in this work and therefore we can ignore numerical errors and drop the superscript $(\infty)$ henceforth.

\subsection{Comparing the second-order stochastic approach to the OD limit and perturbative QFT}

We now have a non-perturbative approach for finding the eigenvalues associated with the second-order Fokker-Planck equation (\ref{phi-pi_fokker-planck_eq}). The only perturbative effect in these eigenvalues is in the relationship between the parameters of the second-order stochastic approach and QFT. Thus, we have the spectrum of solutions that covers the region $\frac{\lambda H^2}{m^2}\ll 1$ in the parameter space (the orange region in Fig. \ref{fig:model_limitations_comparison}). This encompasses the parameter space where the OD approximation is valid. However, thus far we have not explicitly checked that the OD and second-order results agree. To do this, we calculate the first non-zero eigenvalue in each method in the region where the OD approach is valid, $m^2/H^2\ll 1$ and $\lambda\ll m^2/H^2$. 

We will also do an explicit comparison with the perturbative QFT eigenvalue (\ref{QFT_leading eigenvalue}). This is done to check that the eigenvalues of the second-order stochastic approach behave as expected; there should be a good level agreement in the regimes where the established approximations are valid and a degree of difference when they are not.

\subsubsection{Example 1: $m^2/H^2=0.1$}

\begin{figure}
    \includegraphics[width=140mm]{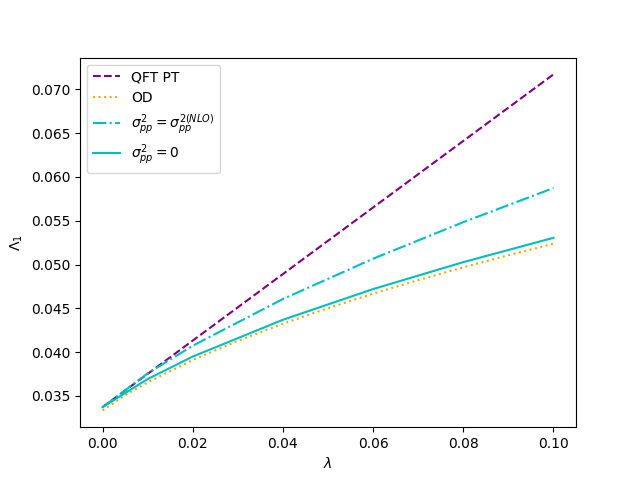}
    \caption{A plot of the first excited eigenvalue $\Lambda_1$ as a function of $\lambda$ for $m^2/H^2=0.1$ using perturbative QFT (purple, dashed), OD stochastic (orange, dotted) and second-order stochastic (cyan) approaches, with $\sigma_{pp}^2=\sigma_{pp}^{2(NLO)}$ (dot-dashed) and $\sigma_{pp}^2
    =0$ (solid).}
    \label{fig:m20.1_match_v_OD_v_QFT}
\end{figure}

We will start by considering two examples of how $\Lambda_1$ compares to the equivalent quantities in the other two approximations by plotting them as a function of $\lambda$ for constant $m^2/H^2$. For clarity, we will label the first non-zero state of the second-order stochastic approach as $\Lambda_1^{(SO)}$. In both Fig. \ref{fig:m20.1_match_v_OD_v_QFT} and \ref{fig:m20.3_match_v_OD_v_QFT}, the solid cyan line shows the choice $\sigma_{pp}^2=0$ (\ref{matched_sigma_pp}) with the choice $\sigma_{pp}^{2(NLO)}$ (Eq. (59c) of Ref. \cite{Cable:2021}) shown for comparison as the dot-dashed cyan line. One can see that in both figures, the two cyan lines diverge from one another as $\lambda$ increases, suggesting that the choice of $\sigma_{pp}^2$ is important. This is not the case. The reason for the discrepancy is because we have not included the one-loop corrections to the stochastic parameters, which enter at relative order $\frac{\lambda H^2}{m^2}$, so this difference really tells us the size of such corrections. If one were to include these, the choice of $\sigma_{pp}^2$ would be irrelevant. This will be undertaken in future work \cite{in_preparation}.

The first example is for $m^2/H^2=0.1$. This is chosen because the mass is sufficiently small such that the OD stochastic approach will be valid beyond both the perturbative QFT and the second-order stochastic approach. Consider Fig. \ref{fig:m20.1_match_v_OD_v_QFT}. One can see that for small $\lambda$, all three models converge. This is as expected because it is in this limit that all three models are valid. As we move towards higher $\lambda$, $\frac{\lambda H^4}{m^4}$ quickly becomes comparable to 1 and therefore the perturbative QFT eigenvalue diverges from the other three curves. This divergence is large, which is no surprise because even at $\lambda=0.01$, $\frac{\lambda H^4}{m^4}=1$ so we are already out of the regime of validity for perturbative QFT. One can see that there is good agreement between the OD and second-order stochastic approaches throughout for $\sigma_{pp}^2=0$, as expected as we are still in the region where the two models should agree.

\subsubsection{Example 2: $m^2/H^2=0.3$}

\begin{figure}
    \includegraphics[width=140mm]{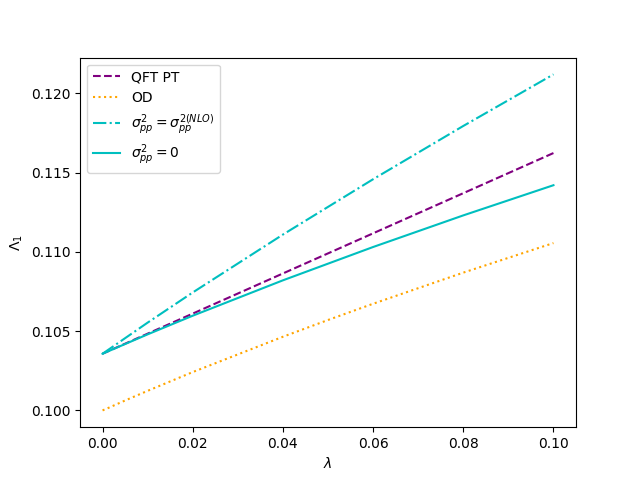}
    \caption{A plot of the first excited eigenvalue $\Lambda_1$ as a function of $\lambda$ for $m^2/H^2=0.3$ using perturbative QFT (purple, dashed), OD stochastic (orange, dotted) and second-order stochastic (cyan) approaches, with $\sigma_{pp}^2=\sigma_{pp}^{2(NLO)}$ (dot-dashed) and $\sigma_{pp}^2
    =0$ (solid).}
    \label{fig:m20.3_match_v_OD_v_QFT} 
\end{figure}

For our second example, we will consider a larger mass $m^2/H^2=0.3$ such that the OD stochastic results become less reliable. Consider Fig. \ref{fig:m20.3_match_v_OD_v_QFT}. One can see that for small $\lambda$ the second-order stochastic and perturbative QFT results agree well but as one increases $\lambda$ the two results diverge from each other. This is once again because increasing $\lambda$ results in an increase of $\frac{\lambda H^4}{m^4}$. Conversely, even at small $\lambda$, the OD stochastic approach gives a different value for the eigenvalue compared to the other two approaches, suggesting that even at $m^2/H^2=0.3$ we are at too high a mass for the OD stochastic approach to be trustworthy.

\subsubsection{OD v second-order stochastic approaches}
\label{subsubsec:OD_v_matching_comparison}

From these two examples, we can see that the behaviour of the second-order stochastic approach is as expected; there is agreement and difference in the regimes where one would expect to find them. To make this more quantitative, we will now consider more carefully the difference between eigenvalues between the second-order stochastic and other two approaches.

\begin{figure}[ht]
    \centering
    \includegraphics[width=150mm]{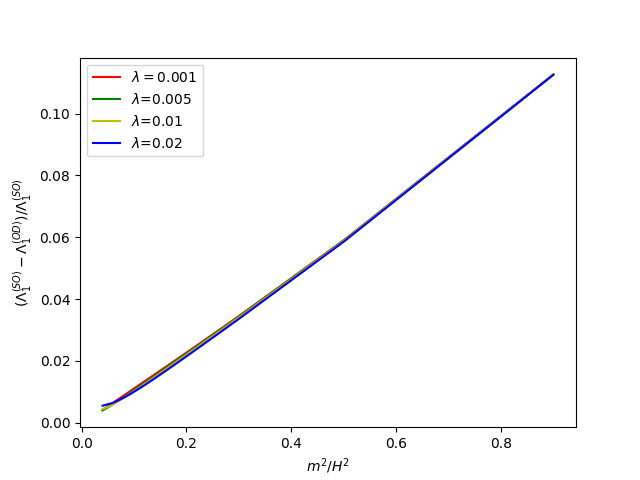}
    \caption{A plot of $\frac{\Lambda_1^{(SO)}-\Lambda_1^{(OD)}}{\Lambda_1^{(SO)}}$ against $m^2/H^2$ for $\lambda=0.001$ (red), $0.005$ (green), $0.01$ (yellow), $0.02$ (blue).}
    \label{fig:eval_v_m2_OD_v_match_diff}
\end{figure}

First, we will consider the difference between the second-order and OD stochastic results. We take the relative difference between the second-order and OD eigenvalues $\frac{\Lambda_1^{(SO)}-\Lambda_1^{(OD)}}{\Lambda_1^{(SO)}}$ as a function of $m^2/H^2$. The use of this scale is so that as one increases $m^2/H^2$ the OD stochastic approach becomes less reliable so we expect to see a difference between the two results. In Fig. \ref{fig:eval_v_m2_OD_v_match_diff}, we have plotted the relative difference for different values of $\lambda$ for the case when $\sigma_{pp}^2=0$. We immediately see that all the curves follow the same linearly increasing behaviour. As we increase $m^2/H^2$ to the right of the figure, we see that the relative difference increases as expected.

\subsubsection{Perturbative QFT v second-order stochastic approaches}
\label{subsubsec:QFT_v_matching_comparison}

We will now do the same analysis with a comparison of the second-order stochastic and perturbative QFT eigenvalues where we plot the eigenvalue difference $\frac{\Lambda_1^{(QFT)}-\Lambda_1^{(SO)}}{\Lambda_1^{(SO)}}$ for several values of $\lambda$ (solid lines in Fig. \ref{fig:eval_v_m2_QFT_v_OD_diff}). The difference is that we will now use $\frac{\lambda H^4}{m^4}$ on the $x$-axis since this is the parameter where we will see a breakdown of perturbative QFT. We see the expected behaviour; for small $\frac{\lambda H^4}{m^4}$, all the curves converge to 0. As one increase, $\frac{\lambda H^4}{m^4}$, we see an increasing relative difference between the two eigenvalues due to a breakdown of the perturbative QFT.

\subsubsection{OD stochastic v perturbative QFT approaches}
\label{subsubsec:OD_v_QFT_comparison}

\begin{figure}
    \centering
    \includegraphics[width=150mm]{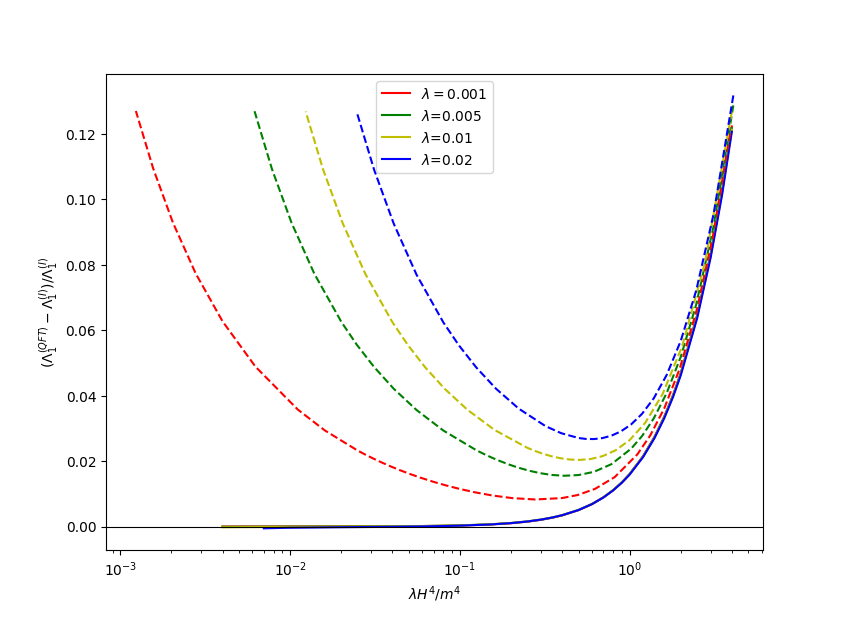}
    \caption{A plot of $\frac{\Lambda_1^{(QFT)}-\Lambda_1^{(I)}}{\Lambda_1^{(I)}}$, where $I\in\{\text{SO},\text{OD}\}$ as a function of $\frac{\lambda H^4}{m^4}$ for $\lambda=0.001$ (red), $\lambda=0.005$ (green), $\lambda=0.01$ (yellow) and $\lambda=0.02$ (blue). The solid lines show the relative difference between the SO and QFT eigenvalues, which lie on directly on top of each other for most $\frac{\lambda H^4}{m^4}$ values, while the dashed lines show the relative difference between the OD and QFT eigenvalues.}
    \label{fig:eval_v_m2_QFT_v_OD_diff}
\end{figure}

The final comparison we will make is between the two established approximations; perturbative QFT and the OD stochastic approach. The dotted lines in Fig. \ref{fig:eval_v_m2_QFT_v_OD_diff} plots the eigenvalue difference between the two approaches, $\frac{\Lambda_1^{(QFT)}-\Lambda_1^{(OD)}}{\Lambda_1^{(OD)}}$ as a function of $\frac{\lambda H^4}{m^4}$ for the four $\lambda$ values. 

On the right hand side of the plot, we can see that the deviation from the QFT result follows the same pattern as that of the second-order stochastic approach. This is unsurprising because, as we move to higher $\frac{\lambda H^4}{m^4}$, we are moving to smaller $m^2/H^2$, the limit where the OD and second-order stochastic approaches agree. In this regime, perturbative QFT is breaking down so we see a high relative difference between it and the stochastic approaches. As we move to smaller values of $\frac{\lambda H^4}{m^4}$, the dotted curves dip to some minimum before turning upward. As one moves left, there is an increasing relative difference between the two; this is now due to the breakdown of the OD stochastic approach since we are getting to high $m^2/H^2$ values. One can see that the second-order stochastic approach continues towards a zero relative difference, indicating the region where the OD approach breaks down but the other two approximations are still valid.

\section{Concluding remarks}
\label{sec:discussion}

We have shown that the second-order stochastic effective theory can be used to calculate correlation functions for self-interacting scalar fields in de Sitter spacetime. The stochastic parameters were determined by matching stochastic correlators to their counterparts in perturbative QFT and a novel numerical calculation was implemented in order to perform computations for fields of mass $m\lesssim H$ and coupling $\lambda\ll m^2/H^2$. This goes beyond the regimes of the established approximations of perturbative QFT and the overdamped stochastic approach, which are limited to $\lambda\ll m^4/H^4$ and $m\ll H$, $\lambda\ll m^2/H^2$ respectively. 
It would be interesting to compare our results to other non-perturbative approaches, but that is beyond the scope of this paper.

Future work is in progress to extend the second-order stochastic approach further to incorporate the full one-loop correction, which will capture the relative order $\mathcal{O}\qty(\frac{\lambda H^2}{m^2})$ contributions. This will improve the results of the current paper and extend the regime of validity of the second-order stochastic theory even further. Ideally, one would like to derive the stochastic parameters from an underlying quantum theory non-perturbatively as opposed to using the perturbative matching procedure; however, it is not clear how one should proceed in this direction.

Regardless, the second-order stochastic effective theory has strong computational power that will be useful in a range of topics in inflationary cosmology, such as the precision calculation of the primordial curvature and isocurvature perturbations in scenarios with light scalar fields. The formalism outlined in this work suggests the stochastic approach has applications beyond its widely-used overdamped state and that it is a method that warrants further study.

\acknowledgements A.C. was supported by a UK Science and Technology Facility Council studentship. A.R. was supported by STFC grants ST/P000762/1 and ST/T000791/1 and IPPP Associateship. A.C. would like to thank Anders Tranberg and Magdalena Eriksson for useful discussions. We would also like to thank Grigoris Pavliotis for his insight.

\appendix

\section{The choice of the $pp$ noise amplitude}
\label{app:pp_noise_amplitude}

The choice of setting $\sigma_{pp}^2=0$ as opposed to fitting it with the subleading contribution is a deviation from our work in Ref. \cite{Cable:2021} and so it is worth discussing here where any difference may occur. The key difference we need to address here is the comparison we made in Eq. (71) of \cite{Cable:2021}. In that work, we had an alternative second-order stochastic approach, where one introduced a cut-off between the sub- and super-horizon modes. We compared this with the matching procedure for free fields and showed that the two models would only reproduce equivalent leading order contributions in the limit $m\ll H$. Here, we show these results again for free fields but also include the matching prescription when $\sigma_{pp}^2=0$ as well. The result is

\begin{subequations}
\label{massless_noise_comparison}
\begin{align}
        \label{qq_massless_noise_comparison}
        \sigma_{qq}^{2(0)}\eval_{m\ll H}&=\frac{H^3}{4\pi^2}, &   \sigma_{cut,qq}^2\eval_{m\ll H}&=\frac{H^3}{4\pi^2}(1+\frac{\epsilon^2}
        {3}+\frac{\epsilon^4}{9});\\
        \label{qp_massless_noise_comparison}
        \sigma_{qp}^{2(0)}\eval_{m\ll H}&=0, & \sigma_{cut,qp}^2\eval_{m\ll H}&=\frac{H^4}{4\pi^2}\qty(-\epsilon^2+\frac{\epsilon^4}{3});\\
        \label{pp_massless_noise_comparison}
        \sigma_{pp}^{2(0)}\eval_{m\ll H}&=\begin{cases}\sigma_{pp}^{2(NLO)}\eval_{m\ll H}=\frac{36H^5}{\pi^2},\\0,\end{cases} & \sigma_{cut,pp}^2\eval_{m\ll H}&=\frac{H^5}{4\pi^2}\epsilon^4.
\end{align}
\end{subequations}

\noindent $\epsilon$ is the cut-off parameter which one takes to be small such that $\epsilon^2\sim0$. We see that the two approaches agree for all three noise amplitudes if $\sigma_{pp}^{2(0)}=0$. This is not what is found if one matches $\sigma_{pp}^2$ with the subleading contribution, when $\sigma_{pp}^{2(0)}=\sigma_{pp}^{2(NLO)}$.

This choice also proves convenient because it is now straightforward to recover the OD stochastic equations from the full second-order theory. Writing the $(\phi,\pi)$ noise amplitudes found through matching in the limit $m\ll H$

\begin{subequations}
    \label{(phi,pi)_noise_sigmapp=0_light_fields}
    \begin{align}
    \sigma_{\phi\phi}^{2}\eval_{m\ll H}&=\frac{H^3}{4\pi^2},\\
    \sigma_{\phi\pi}^{2}\eval_{m\ll H}&=0,\\
    \sigma_{\pi\pi}^{2}\eval_{m\ll H}&=0,
    \end{align}\\
\end{subequations}

\noindent we see that the only non-zero component to the noise amplitude is $\sigma_{\phi\phi}^2$. Considering Eq. (\ref{phi_pi_langevin}) in the OD limit $\Dot{\pi}\ll 3H\pi$, it becomes

\begin{equation}
    \label{langevin_eq_OD_limit}
    \Dot{\phi}+\frac{V'(\phi)}{3H}=\xi_{\phi},
\end{equation}

\noindent which is just the OD stochastic equation (\ref{OD_Langevin}).

\section{Stochastic perturbation theory using general noise}
\label{app:perturbation_theory_general_noise}

In this appendix, we outline the derivation of the stochastic field correlator for a general noise contribution. We include both the timelike and spacelike correlators for completeness.

From Eq. (\ref{O(lambda)_eigenvalues&eigenstates}), the lowest two eigenvalues at $\mathcal{O}(\lambda)$ are

\begin{subequations}
    \label{lowest_3_O(lambda)_eigenvalues_gen_noise}
    \begin{align}
        \Lambda_{00}^{(1)}&=0,\\
         \Lambda_{01}^{(1)}&=\frac{3\alpha\sigma_{pp}^2-4H\alpha\beta^2\sigma_{qp}^2+3H^2\beta^3\sigma_{qq}^2}{8\nu^2H^4\alpha\beta^2}.
    \end{align}
\end{subequations}

\noindent By using Eq. (\ref{O(lambda)_timelike_stochastic_correlator}) and (\ref{O(lambda)_spacelike_stochastic_correlator}) respectively, we can find the timelike and spacelike $q-q$, $q-p$, $p-q$ and $p-p$ stochastic correlators. Including the perturbed noise (\ref{noise_amplitude_perturbative_expansion}), the timelike $(q,p)$ stochastic correlators to $\mathcal{O}(\lambda)$ are 

\begin{footnotesize}
    \begin{subequations}
    \label{(q,p)_timelike_stochastic_correlators_gen_noise}
    \begin{align}
        \label{qq_corr_gen_noise}
        \begin{split}
            \expval{q(0,\mathbf{x})q(t,\mathbf{x})}=&\qty[\frac{\sigma_{qq}^{2(0)}}{2H\alpha}+\lambda\qty(\frac{\sigma_{qq}^{2(1)}}{2H\alpha}+\frac{\qty(3\nu H\sigma_{qq}^{2(0)}-\alpha\sigma_{qp}^{2(0)})\qty(3\alpha\sigma_{pp}^{2(0)}-4H\alpha\beta^2\sigma_{qp}^{2(0)}+3H^2\beta^3\sigma_{qq}^{2(0)})}{48\nu^3H^7\alpha^3\beta^2})]\\&\times e^{-\qty(\alpha H+\lambda\frac{3\alpha\sigma_{pp}^{2(0)}-4H\alpha\beta^2\sigma_{qp}^{2(0)}+3H^2\beta^3\sigma_{qq}^{2(0)}}{8\nu^2H^4\alpha\beta^2}) t}\\
            &+ \qty[\lambda\qty(\frac{\sigma_{qp}^{2(0)}\qty(-3\alpha \sigma_{pp}^{2(0)}+4H\alpha \beta^2\sigma_{qp}^{2(0)}-3H^2\beta^3\sigma_{qq}^{2(0)})}{48\nu^3H^7\alpha\beta^3})]e^{-\qty(\beta H-\lambda\frac{3\alpha\sigma_{pp}^{2(0)}-4H\alpha\beta^2\sigma_{qp}^{2(0)}+3H^2\beta^3\sigma_{qq}^{2(0)}}{8\nu^2H^4\alpha\beta^2})t},\\&
        \end{split}\\
        \label{pq_corr_gen_noise}
        \begin{split}
            \expval{p(0,\mathbf{x})q(t,\mathbf{x})}=&\qty[\frac{\sigma_{qp}^{2(0)}}{3H}+\lambda\qty(\frac{\sigma_{qp}^{2(1)}}{3H}+\frac{\qty(\nu H^2\beta^2\sigma_{qq}^{2(0)}-\alpha \sigma_{pp}^{2(0)})\qty(3\alpha\sigma_{pp}^{2(0)}-4H\alpha\beta^2\sigma_{qp}^{2(0)}+3H^2\beta^3\sigma_{qq}^{2(0)})}{48\nu^3H^7\alpha^2\beta^3})]\\&\times e^{-\qty(\alpha H+\lambda\frac{3\alpha\sigma_{pp}^{2(0)}-4H\alpha\beta^2\sigma_{qp}^{2(0)}+3H^2\beta^3\sigma_{qq}^{2(0)}}{8\nu^2H^4\alpha\beta^2}) t}
            \\&
            +\qty[\lambda\qty(\frac{\sigma_{pp}^{2(0)}(-3\alpha\sigma_{pp}^{2(0)}+4H\alpha\beta^2\sigma_{qp}^{2(0)}-3H^2\beta^3\sigma_{qq}^{2(0)})}{32\nu^3H^7\alpha\beta^4})]e^{-\qty(\beta H-\lambda\frac{3\alpha\sigma_{pp}^{2(0)}-4H\alpha\beta^2\sigma_{qp}^{2(0)}+3H^2\beta^3\sigma_{qq}^{2(0)}}{8\nu^2H^4\alpha\beta^2})t},\\&
        \end{split}\\
        \label{qp_corr_gen_noise}
        \begin{split}
            \expval{q(0,\mathbf{x})p(t,\mathbf{x})}=&\qty[\lambda\qty(\frac{\sigma_{qq}^{(0)}\qty(3\alpha\sigma_{pp}^{2(0)}-4H\alpha\beta^2\sigma_{qp}^{2(0)}+3H^2\beta^3\sigma_{qq}^{2(0)})}{32\nu^3H^5\alpha^2\beta})]e^{-\qty(\alpha H+\lambda\frac{3\alpha\sigma_{pp}^{2(0)}-4H\alpha\beta^2\sigma_{qp}^{2(0)}+3H^2\beta^3\sigma_{qq}^{2(0)}}{8\nu^2H^4\alpha\beta^2}) t}
            \\&
            +\qty[\frac{\sigma_{qp}^{2(0)}}{3H}+\lambda\qty(\frac{\sigma_{qp}^{2(1)}}{3H}+\frac{\qty(\nu\sigma_{pp}^{2(0)}+H^2\beta^3\sigma_{qq}^{2(0)})\qty(3\alpha\sigma_{pp}^{2(0)}-4H\alpha\beta^2\sigma_{qp}^{2(0)}+3H^2\beta^3\sigma_{qq}^{2(0)})}{48\nu^3H^7\alpha\beta^4})]\\&\times e^{-\qty(\beta H-\lambda\frac{3\alpha\sigma_{pp}^{2(0)}-4H\alpha\beta^2\sigma_{qp}^{2(0)}+3H^2\beta^3\sigma_{qq}^{2(0)}}{8\nu^2H^4\alpha\beta^2})t},
            \\&
        \end{split}\\
        \begin{split}
            \label{pp_corr_gen_noise}
            \expval{p(0,\mathbf{x})p(t,\mathbf{x})}=&\qty[\lambda\qty(\frac{\sigma_{qp}^{2(0)}\qty(-3\alpha\sigma_{pp}^{2(0)}+4H\alpha\beta^2\sigma_{qp}^{2(0)}-3H^2\beta^3\sigma_{qq}^{2(0)})}{48\nu^3H^5\alpha\beta})]e^{-\qty(\alpha H+\lambda\frac{3\alpha\sigma_{pp}^{2(0)}-4H\alpha\beta^2\sigma_{qp}^{2(0)}+3H^2\beta^3\sigma_{qq}^{2(0)}}{8\nu^2H^4\alpha\beta^2}) t}\\&
            +\qty[\frac{\sigma_{pp}^{2(0)}}{2H\beta}+\lambda\qty(\frac{\sigma_{pp}^{2(1)}}{2H\beta}+\frac{\qty(3\nu\sigma_{pp}^{2(0)}+H\beta^2\alpha\sigma_{qp}^{2(0)})\qty(3\alpha\sigma_{pp}^{2(0)}-4H\alpha\beta^2\sigma_{qp}^{2(0)}+3H^2\beta^3\sigma_{qq}^{2(0)})}{48\nu^3H^6\alpha\beta^4})]\\&\times e^{-\qty(\beta H-\lambda\frac{3\alpha\sigma_{pp}^{2(0)}-4H\alpha\beta^2\sigma_{qp}^{2(0)}+3H^2\beta^3\sigma_{qq}^{2(0)}}{8\nu^2H^4\alpha\beta^2})t}
        \end{split}
    \end{align}
    \end{subequations}
\end{footnotesize}

\noindent and their spacelike counterparts are

\begin{footnotesize}
\begin{subequations}
    \label{(q,p)_spacelike_stochastic_correlators_gen_noise}
    \begin{align}
    \label{spacelike_qq_gen_noise}
        \begin{split}
            \expval{q(t,\mathbf{0})q(t,\mathbf{x})}=&\qty[\frac{\sigma_{qq}^{2(0)}}{2H\alpha}+\lambda\qty(\frac{\sigma_{qq}^{2(1)}}{2H\alpha}+\frac{\qty(3\alpha\sigma_{pp}^{2(0)}-4H\alpha\beta^2\sigma_{qp}^{2(0)}+3H^2\beta^3\sigma_{qq}^{2(0)})\qty(3\alpha\sigma_{qp}^{2(0)}-3\nu H\beta\sigma_{qq}^{2(0)})}{48\nu^3H^7\alpha^3\beta^3})]\\&\times \abs{Ha(t)\mathbf{x}}^{-2\alpha-\lambda\frac{3\alpha\sigma_{pp}^{2(0)}-4H\alpha\beta^2\sigma_{qp}^{2(0)}+3H^2\beta^3\sigma_{qq}^{2(0)}}{4\nu^2H^5\alpha\beta^2}}\\&
            +\qty[\lambda\qty(\frac{\sigma_{qp}^{2(0)}(-3\alpha\sigma_{pp}^{2(0)}+4H\alpha\beta^2\sigma_{qp}^{2(0)}-3H^2\beta^3\sigma_{qq}^{2(0)})}{24\nu^3H^7\alpha\beta^3})]\abs{Ha(t)\mathbf{x}}^{-3},
        \end{split}\\
    \label{spacelike_qp_gen_noise}
        \begin{split}
            \expval{q(t,\mathbf{0})p(t,\mathbf{x})}=&\expval{p(t,\mathbf{0})q(t,\mathbf{x})}=\\
            &\qty[\lambda\qty(\frac{\sigma_{qq}^{2(0)}\qty(-3\alpha\sigma_{pp}^{2(0)}+4H\alpha\beta^2\sigma_{qp}^{2(0)}-3H^2\beta^3\sigma_{qq}^{2(0)})}{32\nu^3H^5\alpha^2\beta})]\abs{Ha(t)\mathbf{x}}^{-2\alpha-\lambda\frac{3\alpha\sigma_{pp}^{2(0)}-4H\alpha\beta^2\sigma_{qp}^{2(0)}+3H^2\beta^3\sigma_{qq}^{2(0)}}{4\nu^2H^5\alpha\beta^2}}\\&
            +\qty[\lambda\qty(\frac{\sigma_{pp}^{2(0)}\qty(-3\alpha\sigma_{pp}^{2(0)}+4H\alpha\beta^2\sigma_{qp}^{2(0)}-3H^2\beta^3\sigma_{qq}^{2(0)})}{32\nu^3H^7\alpha\beta^4})]\abs{Ha(t)\mathbf{x}}^{-2\beta+\lambda\frac{3\alpha\sigma_{pp}^{2(0)}-4H\alpha\beta^2\sigma_{qp}^{2(0)}+3H^2\beta^3\sigma_{qq}^{2(0)}}{4\nu^2H^5\alpha\beta^2}}\\&
            +\qty[\frac{\sigma_{qp}^{2(0)}}{3H}+\lambda\qty(\frac{\sigma_{qp}^{2(1)}}{3H}+\frac{\qty(\sigma_{pp}^{2(0)}+H^2\beta^2\sigma_{qq}^{2(0)})\qty(3\alpha\sigma_{pp}^{2(0)}-4H\alpha\beta^2\sigma_{qp}^{2(0)}+3H^2\beta^3\sigma_{qq}^{2(0)})}{48\nu^3H^7\alpha\beta^3})]\abs{Ha(t)\mathbf{x}}^{-3},
        \end{split}\\
    \label{spacelike_pp_gen_noise}
        \begin{split}
            \expval{p(t,\mathbf{0})p(t,\mathbf{x})}=&\qty[\frac{\sigma_{pp}^{2(0)}}{2H\beta}+\lambda\qty(\frac{\sigma_{pp}^{2(1)}}{2H\beta}+\frac{\qty(3\nu\sigma_{pp}^{2(0)}+3H\beta^2\sigma_{qp}^{2(0)})\qty(3\alpha\sigma_{pp}^{2(0)}-4H\alpha\beta^2\sigma_{qp}^{2(0)}+3H^2\beta^3\sigma_{qq}^{2(0)}))}{48\nu^3H^6\alpha\beta^4})]\\&\times \abs{Ha(t)\mathbf{x}}^{-2\beta+\lambda\frac{3\alpha\sigma_{pp}^{2(0)}-4H\alpha\beta^2\sigma_{qp}^{2(0)}+3H^2\beta^3\sigma_{qq}^{2(0)}}{4\nu^2H^5\alpha\beta^2}}
            \\&\qty[\lambda\qty(\frac{\sigma_{qp}^{2(0)}(-3\alpha\sigma_{pp}^{2(0)}+4H\alpha\beta^2\sigma_{qp}^{2(0)}-3H^2\beta^3\sigma_{qq}^{2(0)}))}{24\nu^3H^5\alpha\beta})]\abs{Ha(t)\mathbf{x}}^{-3}.
        \end{split}
    \end{align}
\end{subequations}
\end{footnotesize}

\noindent Substituting these expressions into Eq. (\ref{phi-phi_pq}), we obtain an expression for the $\phi-\phi$ stochastic correlator up to first order in $\lambda$. The timelike version is

\begin{small}
\begin{equation}
    \label{phi-phi_timelike_stochastic_correlator_gen_noise}
    \begin{split}
        \expval{\phi(0,\mathbf{x})\phi(t,\mathbf{x})}=&\frac{1}{1-\frac{\alpha}{\beta}}\Bigg[\frac{\sigma_{qq}^{2(0)}}{2H\alpha }-\frac{\sigma_{qp}^{2(0)}}{3H^2\beta}+\lambda\Bigg(\frac{\sigma_{qq}^{2(1)}}{2H\alpha}-\frac{\sigma_{qp}^{2(1)}}{3H^2\beta}\\&+\Bigg(2H\alpha^4\beta\sigma_{qp}^{2(0)}\qty(-3\sigma_{pp}^{2(0)}+4H\beta^2\sigma_{qp}^{2(0)})\\&+9H^4\beta^7(\sigma_{qq}^{2(0)})^2+9H^2\alpha\beta^4\sigma_{qq}^{2(0)}\qty(-\sigma_{pp}^{2(0)}+2H\beta^2\qty(\sigma_{qp}^{2(0)}+H\sigma_{qq}^{2(0)}))\\&-3\alpha^3\qty(6(\sigma_{pp}^{2(0)})^2+6H^3\beta^4\sigma_{qp}^{2(0)}\sigma_{qq}^{2(0)}-H\beta^2\sigma_{pp}^{2(0)}\qty(8\sigma_{qp}^{2(0)}+3H\sigma_{qq}^{2(0)}))\\&+H\alpha^2\beta^3\qty(6\sigma_{pp}^{2(0)}\sigma_{qp}^{2(0)}+H\beta^2\qty(-8(\sigma_{qp}^{2(0)})^2-24H\sigma_{qp}^{2(0)}\sigma_{qq}^{2(0)}+9H^2(\sigma_{qq}^{2(0)})^2))\Bigg)/\qty(288\nu^3H^8\alpha^3\beta^4)\Bigg)\Bigg]\\&\times e^{-\qty(\alpha H+\lambda\frac{3\alpha\sigma_{pp}^{2(0)}-4H\alpha\beta^2\sigma_{qp}^{2(0)}+3H^2\beta^3\sigma_{qq}^{2(0)}}{8\nu^2H^4\alpha\beta^2}) t}
        \\&+\\&
        \frac{1}{1-\frac{\alpha}{\beta}}\Bigg[\frac{\sigma_{pp}^{2(0)}}{2H^3\beta^3}-\frac{\sigma_{qp}^{2(0)}}{3H^2\beta}+\lambda\Bigg(\frac{\sigma_{pp}^{2(1)}}{2H^3\beta^3}-\frac{\sigma_{qp}^{2(1)}}{3H^2\beta}
        \\&+\Bigg(9\alpha\qty(-\alpha^2+2\alpha \beta+\beta^2)(\sigma_{pp}^{2(0)})^2+6H\alpha \beta^2\qty(3\alpha^2-4\alpha\beta-3\beta^2)\sigma_{qp}^{2(0)}\sigma_{pp}^{2(0)}\\&+6H^3\beta^5\qty(\alpha^2+4\alpha\beta-\beta^2)\sigma_{qp}^{2(0)}\sigma_{qq}^{2(0)}-18H^4\beta^7(\sigma_{qq}^{2(0)})^2\\&+H^2\beta^3\qty(\beta^2-\alpha^2)\qty(8\alpha\beta(\sigma_{qp}^{2(0)})^2+9\sigma_{pp}^{2(0)}\sigma_{qq}^{2(0)}))\Bigg)/\qty(288\nu^3H^8\alpha\beta^6)\Bigg)\Bigg]
        \\&\times e^{-\qty(\beta H-\lambda\frac{3\alpha\sigma_{pp}^{2(0)}-4H\alpha\beta^2\sigma_{qp}^{2(0)}+3H^2\beta^3\sigma_{qq}^{2(0)}}{8\nu^2H^4\alpha\beta^2})t}
    \end{split}
\end{equation}
\end{small}

\noindent while the spacelike version is\\

\begin{footnotesize}
\begin{equation}
    \label{phi-phi_spacelike_stochastic_correlator_gen_noise}
    \begin{split}
        \expval{\phi(t,\mathbf{0})\phi(t,\mathbf{x})}=&\frac{1}{1-\frac{\alpha}{\beta}}\qty[\frac{\sigma_{qq}^{2(0)}}{2H\alpha}+\lambda\qty(\frac{\sigma_{qq}^{2(1)}}{2H\alpha}+\frac{\qty(H\beta(3\alpha-\beta)\sigma_{qq}^{2(0)}+2\alpha\sigma_{qp}^{2(0)})\qty(3\alpha\sigma_{pp}^{2(0)}-4H\alpha\beta^2\sigma_{qp}^{2(0)}+3H^2\beta^3\sigma_{qq}^{2(0)})}{32\nu^3H^7\alpha^3\beta^3})]\\&\times \abs{Ha(t)\mathbf{x}}^{-2\alpha-\lambda\frac{3\alpha\sigma_{pp}^{2(0)}-4H\alpha\beta^2\sigma_{qp}^{2(0)}+3H^2\beta^3\sigma_{qq}^{2(0)}}{4\nu^2H^5\alpha\beta^2}}\\&
        +\\&
        \frac{1}{1-\frac{\alpha}{\beta}}\qty[\frac{\sigma_{pp}^{2(0)}}{2H^3\beta^3}+\lambda\qty(\frac{\sigma_{pp}^{2(1)}}{2H^3\beta^3}+\frac{(\qty(3\beta-\alpha)\sigma_{pp}^{2(0)}+2H\beta^2\sigma_{qp}^{2(0)})(3\alpha\sigma_{pp}^{2(0)}-4H\alpha\beta^2\sigma_{qp}^{2(0)}+3H^2\beta^3\sigma_{qq}^{2(0)})}{32\nu^3H^8\alpha\beta^6})]\\&\times \abs{Ha(t)\mathbf{x}}^{-2\beta+\lambda\frac{3\alpha\sigma_{pp}^{2(0)}-4H\alpha\beta^2\sigma_{qp}^{2(0)}+3H^2\beta^3\sigma_{qq}^{2(0)}}{4\nu^2H^5\alpha\beta^2}}\\&
        +\\&
        \frac{1}{1-\frac{\alpha}{\beta}}\Bigg[-\frac{2\sigma_{qp}^{2(0)}}{3H^2\beta}+\lambda\Bigg(-\frac{2\sigma_{qp}^{2(1)}}{3H^2\beta}\\&+\Bigg(2H\alpha^2\beta\sigma_{qp}^{2(0)}\qty(4H\beta^2\sigma_{qp}^{2(0)}-3\sigma_{pp}^{2(0)})-3H^2\beta^3\sigma_{qq}^{2(0)}\qty(3\sigma_{pp}^{2(0)}+H\beta^2\qty(2\sigma_{qp}^{2(0)}+3H\sigma_{qq}^{2(0)}))\\&-\alpha\qty(9(\sigma_{pp}^{2(0)})^2-3H\beta^2\sigma_{pp}^{2(0)}\qty(2\sigma_{qp}^{2(0)}-3H\sigma_{qq}^{2(0)})-2H^2\beta^4\sigma_{qp}^{2(0)}\qty(4\sigma_{qp}^{2(0)}+3H\sigma_{qq}^{2(0)}))\Bigg)/\qty(72\nu^3H^8\alpha\beta^4)\Bigg)\Bigg]\\&\times\abs{Ha(t)\mathbf{x}}^{-3}.
    \end{split}
\end{equation}
\end{footnotesize}

\noindent This reduces to Eq. (\ref{phi-phi_spacelike_stochastic_correlator}) if we use the free noise amplitudes (\ref{matched_(p,q)_noise_free}).

\bibliographystyle{unsrt}
\bibliography{main.bib}

\end{document}